\documentclass[longauth]{aa} 
\usepackage{xcolor}
\usepackage{support-caption}
\usepackage{subcaption}
\usepackage{supertabular,booktabs}
\usepackage{longtable}
\usepackage{graphicx}
\usepackage[flushleft]{threeparttable}

\newcommand{\hii}{\textrm{H}\textsc{ii}}

\newcommand{\oiii}{[\textrm{O}\textsc{iii}]}
\newcommand{\oii}{[\textrm{O}\textsc{ii}]}

\newcommand{\oiilam}{[\textrm{O}\textsc{ii}]\ensuremath{\lambda3727}}

\newcommand{\oiiialone}{[\textrm{O}\textsc{iii}]}

\newcommand{\oiiidoub}{[\textrm{O}\textsc{iii}]\ensuremath{\lambda\lambda4959,5007}}
\newcommand{\ha}{\ifmmode { H}\alpha \else H$\alpha$\fi}
\newcommand{\hb}{\ifmmode { H}\beta \else H$\beta$\fi}
\newcommand{\lya}{\ifmmode { Ly}\alpha \else Ly$\alpha$\fi}
\newcommand{\pg}{\ifmmode { P}\gamma \else Pa$\gamma$\fi}
\newcommand{\lyb}{\ifmmode { Ly}\beta \else Ly$\beta$\fi}
\newcommand{\lyg}{\ifmmode { Ly}\gamma \else Ly$\gamma$\fi}

\newcommand{\flyc}{\ifmmode \mathrm{f}_\mathrm{ esc}\mathrm{(LyC)} \else $\mathrm{f}_\mathrm{ esc}\mathrm{(LyC)}$\fi}

\def\ergs{\ifmmode \mathrm{erg\hspace{1mm}s}^{-1} \else erg s$^{-1}$\fi}

\def\micron{\ifmmode \mu\mathrm{m} \else $\mu$m\fi}
\def\msun{\ifmmode \mathrm{M}_{\odot} \else M$_{\odot}$\fi}
\def\msunyr{\ifmmode \mathrm{M}_{\odot} \hspace{1mm}{ yr}^{-1} \else $\mathrm{M}_{\odot}$ yr$^{-1}$\fi}
\def\zsun{\ifmmode Z_{\odot} \else Z$_{\odot}$\fi}
\def\lsun{\ifmmode L_{\odot} \else L$_{\odot}$\fi}
\def\mstar{\ifmmode \mathrm{M}_{\star} \else M$_{\star}$\fi}

\usepackage{txfonts}
\newcommand{\orcid}[1]{\href{https://orcid.org/#1}{\includegraphics[width=10pt]{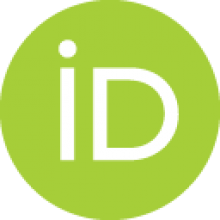}}}
\usepackage{hyperref}
\hypersetup{colorlinks, % = true per default
 linkcolor = blue,
 urlcolor = blue,
 citecolor = blue,
 anchorcolor = blue
}
\begin{document} 

\title{A Lyman continuum analysis for $\sim 100$ galaxies at $z_{\text{spec}} \sim 3$ in the Abell 2744 Cluster Field}
 
 \subtitle{}
 \author{Y. Liu \fnmsep\thanks{E-mail: yuchenliu.pku@gmail.com}
 \inst{1,2,3}
 \and
 S. Mascia 
 \inst{4}
 \and
 L. Pentericci 
 \inst{1}
 \and 
 P. Watson
 \inst{5}
 \and
 A. Alavi
 \inst{12}
\and
 P. Bergamini
 \inst{19}
  \and
 M. Brada\v{c}
 \inst{14,15}
 \and
 A. Calabrò
 \inst{1}
  \and
 K. Glazebrook
 \inst{17}
 \and
 A. Henry
 \inst{13}
  \and
 M. Llerena
 \inst{1}
  \and
 E. Merlin
 \inst{1}
 \and
 B. Metha
 \inst{18}
  \and
 T. Nanayakkara
 \inst{17}
  \and
 L. Napolitano
\inst{1,10}
 \and
 N. Roy
 \inst{6}
 \and 
 B. Siana
 \inst{11}
  \and
 E. Vanzella
 \inst{16}
 \and
 B. Vulcani
 \inst{5}
 \and
  X. Wang
 \inst{7,8,9}
}

 \institute{INAF – Osservatorio Astronomico di Roma, via Frascati 33, 00078, Monteporzio Catone, Italy
 \and 
 Department of Astronomy, School of Physics, Peking University, Beijing 100871,China
 \and
 Kavli Institute for Astronomy and Astrophysics, Peking University, Beijing 100871, China
 \and
 Institute of Science and Technology Austria (ISTA), Am Campus 1, A-3400 Klosterneuburg, Austria 
\and
INAF -- Osservatorio Astronomico di Padova, Vicolo Osservatorio 5, 35122 Padova, Italy
\and
Center for Astrophysical Sciences, William H. Miller III Department of Physics and Astronomy, Johns Hopkins University, Baltimore, MD, 21218
\and
School of Astronomy and Space Science, University of Chinese Academy of Sciences (UCAS), Beijing 100049, China
\and
National Astronomical Observatories, Chinese Academy of Sciences, Beijing 100101, China
\and
Institute for Frontiers in Astronomy and Astrophysics, Beijing Normal University, Beijing 102206, China
\and 
Dipartimento di Fisica, Università di Roma Sapienza, Città Universitaria di Roma - Sapienza, Piazzale Aldo Moro, 2, 00185, Roma, Italy
\and
Department of Physics \& Astronomy, University of California, Riverside, 900 University Avenue, Riverside, CA 92521, USA
\and
IPAC, California Institute of Technology, 1200 E. California Blvd., Pasadena, CA 91125, USA
\and
Space Telescope Science Institute, 3700 San Martin Drive, Baltimore, MD 21218, USA
\and
University of Ljubljana, Faculty of Mathematics and Physics, Jadranska ulica 19, SI-1000 Ljubljana, Slovenia
\and
Department of Physics and Astronomy, University of California Davis, 1 Shields Avenue, Davis, CA 95616, USA
\and
INAF – Osservatorio di Astrofisica e Scienza dello Spazio di
Bologna, via Gobetti 93/3, I-40129, Bologna, Italy
\and
 Centre for Astrophysics and Supercomputing, Swinburne University
of Technology, PO Box 218, Hawthorn, VIC 3122, Australia
\and
School of Physics, The University of Melbourne, VIC 3010, Australia
\and
Dipartimento di Fisica, Università degli Studi di Milano, Via Celoria 16, I-20133 Milano, Italy
}
 \date{Accepted XXX. Received YYY; in original form ZZZ}
% \abstract{}{}{}{}{} 
% 5 {} token are mandatory

 \abstract
 {Identifying Lyman continuum (LyC) leakers at intermediate redshifts is crucial for understanding the properties of cosmic reionizers, as the opacity of the intergalactic medium (IGM) prevents direct detection of LyC emission from sources during the Epoch of Reionization (EoR). In this study, we confirm two new LyC candidate leakers at $z \sim 3$ in the Abell 2744 cluster field, with absolute escape fractions ($f_{\text{esc}}$) of $0.90^{+0.07}_{-0.86}$ and $0.60^{+0.37}_{-0.56}$, respectively. The LyC emission was detected using HST/WFC3/F275W and F336W imaging. These two candidate leakers appear faint ($ M_{\text{UV}} = -18.1 \pm 0.1 \text{ and } -17.81 \pm 0.11$), exhibit blue UV continuum slopes ($\beta = -2.42 \pm 0.05 \text{ and } -1.78 \pm 0.19$), have low masses ($M_\star \sim 10^{7.73} \pm 0.1 \text{ and } 10^{7.07} \pm 0.05 M_\odot$) and show \lya\ equivalent widths of $90 \pm 3$ \AA\ and $28 \pm 12$ \AA, respectively. The discovery of these two LyC candidate leakers was achieved in a catalog of 91 spectroscopically confirmed sources using JWST and/or MUSE public spectra. We also analyze properties that have been proposed as indirect indicators of LyC emission, like \lya, O32 ratio, and $M_\star$: we create subsample of galaxies selected according to such properties, stack the LyC observations of these subsample and assess the limits in escape fractions in the stacks. By analysing the individual candidates and the stacks, in the context of the currently limited sample of known LyC leakers at $z \sim 3$, we aim to enhance our understanding of LyC escape mechanisms and improve our predictions of the LyC $f_{\text{esc}}$ during the EoR.}
 
\maketitle
%-------------------------------------------------------------------

\section{Introduction}
%main topic
The Epoch of Reionization (EoR) began as galaxies increased in number and UV brightness, emitting Lyman continuum (LyC, $\lambda < 912$ \AA) radiation that ionized the surrounding neutral intergalactic medium (IGM). The LyC photons produced by star-forming galaxies can account for the photon budget required to complete reionization only if a substantial fraction of them escapes from the galaxies' interstellar and circumgalactic media (ISM and CGM) into the IGM. Recent surveys by the \textit{Hubble Space Telescope} (HST) and the \textit{James Webb Space Telescope} (JWST) have well-documented the density of star-forming galaxies in the EoR. These studies suggest that star-forming galaxies--rather than AGNs--are likely to dominate the process of cosmic reionization by $z\sim5$ \citep[e.g.,][]{Finkelstein2019, Dayal2020,Bosman2022, dayal2024, jiang2025, whitler2025} and it matches the Thomson optical depth of electron scattering observed in the cosmic microwave background \citep[CMB,][]{Planck2020}.

However, at redshifts $z \geq 4.5$, detecting LyC photons escaping from galaxies is impossible due to the increased IGM opacity. IGM absorption would result in only a 20\% likelihood of detecting LyC leakage from galaxies at $z \sim 4$, decreasing and going to zero at even higher redshifts \citep[][]{inoue2008,inoue2014}.
%The LyC emission can therefore only be detected at low and intermediate redshifts. 
Researchers have thus started to study lower redshift LyC leakers and identify key ISM properties and physical conditions that facilitate LyC photon escape, which can be thus considered as indirect tracers of LyC escape \citep[e.g.,][]{yamanaka2020, Izotov2018b, Marchi2018, Verhamme2017}. Such properties can then be identified also in EoR galaxies, to understand their role in cosmic reionization.

One of the most reliable indicators identified at $z\lesssim 4$ is the presence of strong \lya\ emission \citep[e.g.,][]{Pahl2021, Gazagnes2020}. \cite{Verhamme2017} proposed that a \lya\ profile with a double peak and narrow separation between the peaks might indicate significant LyC leakage, as smaller peak separation corresponds to a lower column density of neutral gas, allowing both the \lya\ and LyC to escape \citep{Rivera-Thorsen2019,Izotov2018b,Vanzella2020, Verhamme2015,Henry2015}. However, at $z>6.5$, \lya\ is attenuated due to its resonant nature and the increasingly neutral IGM as we approach the EoR \citep[][]{Pentericci_2018b, mason2019, jung2020, Ouchi_2020, bolan2021,Napolitano2024}. Therefore this emission line cannot be typically employed to indicate LyC escape at such high redshift \citep[e.g.,][]{Napolitano2024}. 

Another useful indicator is the \oiiidoub/ \oiilam\ (O32) line ratio. \cite{Nakajima2014} first found evidence that high O32 values suggest partially incomplete \hii\ regions, allowing some LyC photons to escape \citep[][]{Jaskot_2013,Nakajima2014, Marchi2018, Izotov2018b}. This correlation is characterized by significant scatter in low-redshift studies \citep[e.g.,][]{Izotov2018b, Nakajima2020, Flury2022}. In practice, while a high O32 flux ratio is necessary for a significant measurement of $f_{\text{esc}}$, it is not sufficient by itself to identify a LyC leaker, as factors such as viewing angles, metallicity, and ionization parameter variations may also play a role \citep[e.g.,][]{Bassett2019, Katz2020}. Beyond these strong-line diagnostics, recent low-redshift studies have explored weak [S\textsc{ii}] emission emerging as a promising new indirect indicator of LyC escape \citep{wang2019,wang2021,Roy2024}.

%Another diagnostic for a high $f_{\text{esc}}$ is the star formation rate surface density ($\Sigma_{\text{SFR}}$): feedback from star formation can create bubbles or chimneys in the host galaxy's ISM, linking a high $\Sigma_{\text{SFR}}$ to a high $f_{\text{esc}}$. This connection is supported by detecting some compact LyC emitters \citep[LCEs, e.g.,][]{vanzella2015,deBarros2016, Schaerer2016,Izotov2018b}, even though compactness may not be a defining characteristic of all of them.

\cite{Chisholm2022} recently proposed that the UV $\beta$ slope serves as a reliable predictor of LyC $f_{\text{esc}}$, with bluer $\beta$ values generally linked to higher $f_{\text{esc}}$. \cite{Choustikov2024} predicted a strong correlation between extremely blue $\beta$ values ($< -2.5$) and elevated specific star formation rates (sSFR), which indicates that the young stellar population is more dominant compared to older stars in the galaxy with steeper $\beta$ slope. While $\beta$ is sensitive to dust, galaxies with very low E(B-V) values also show a good correlation with $f_{\text{esc}}$ at lower and intermediate redshifts \citep{Reddy2016, Saldana-Lopez2022}. Given these considerations, $\beta$ is likely to be a useful indicator of $f_{\text{esc}}$, although it is subject to significant scatter.

The Low redshift Ly Continuum Survey (LzLCS+ ) analyzed the first statistical sample of 66 galaxies at $z=0.2-0.4$ with either a detection or very stringent limit on the LyC emission from HST-COS spectra \citep{Flury2022b, Flury2022, Saldana-Lopez2022}. Having complete spectroscopic and photometric information for all sources, they tested many promising indirect diagnostics to understand how they correlate with LyC emission. 
Using this large low-redshift sample, several authors \citep[e.g.,][]{Chisholm2022, Mascia2023, Mascia2024, Lin2024, Jaskot24a, Jaskot24b} derived multivariate predictions of LyC $f_{\text{esc}}$ based on particular subset of indirect indicators, with the ultimate goal of applying these correlations to galaxies at $z\simeq 6-8$ to determine their contribution to reionization. The major limitation of such an exercise is assuming that the conditions which facilitate LyC escape at $z \sim 0.3$ are the same occurring during the EoR i.e. 10 Gyrs earlier. Moreover, the surveys often do not cover the same bands of the physical parameters as those of high-z galaxies do.

Thus, it is of paramount importance to first test the solidity of the indirect diagnostics in $z\simeq 3-4$ galaxies, where a direct detection of LyC is still possible, as these galaxies are much closer in cosmic time to the EoR (just $\sim$ 1 Gyrs), and likely much better analogs of the cosmic reionizers.
Currently, only a handful of galaxies with $f_{\text{esc}} \geq 0.1$ have been identified in this cosmic epoch ($z\simeq 2-4$) \citep[e.g., Ion2 in][]{Vanzella2015, Vanzella2018, Vanzella2020, Fletcher2019, Ji2020, Marques-Chaves2022} and some of the studies constrained the limits from stacks \citep{grazian2016,Pahl2021,mestric2021}. Recent efforts have continued to expand this sample: \cite{Wang2023} identified five potential LyC leakers using WFC3/F275W imaging. \cite{Liu2023} added five individual LyC leakers with escape fractions raging from 0.4 to 0.8. \cite{Jung2024} placed upper limits on the absolute escape fraction ($f_{\rm esc, abs}$) between 3\% and 15\% for galaxies at $1.3\leq z \leq 3$. 
Despite this progress, the number of confirmed leakers remains small, and crucially, we still lack comprehensive information on their indirect diagnostics. Apart from the \lya\ emission which is strongly linked to the LyC emission also at $z \simeq 3$ \citep[e.g,][]{Izotov2017, Verhamme2017, Marchi2018,Roy2023}, only a limited sources have a complete census of rest-frame optical diagnostics such as \oiii, \oii, \hb\ and \textrm{MgII} emission.
Although JWST's near-IR spectroscopy has recently enabled access to these features for an increasing number of sources \citep{Roy2023, Prieto-Lyon2023, llerena2024}, additional data are still required to establish robust correlations and to calibrate indirect diagnostics at higher redshifts where direct LyC detection is not possible.

The aim of this work is to search and study new intermediate-redshift LyC leakers in the Abell 2744 field, taking advantage both of the presence of deep HST imaging obtained with the WFC3/UVIS F275W and F336W filters, which capture the redshifted LyC emission from galaxies at $z \sim 3$, and a very rich optical/near-IR spectroscopic dataset, ranging from ground-based MUSE observations to JWST spectroscopy. This dataset obtained a wide range coverage of wavelengths and enables us to investigate the correlation between multiple properties and LyC escape at intermediate redshifts.

The paper is organized as follows: in Sec.~\ref{sec:data}, we outline the sample selection process; Sec.~\ref{sec:method} describes the NIRISS extraction of the sources, the measurement of key emission lines, and the direct calculation of $f_{\text{esc}}$ values. In Sec.~\ref{sec:results} , we present the main properties of the two candidate leakers and the upper limits for subsamples of the sources; Sec.~\ref{sec:discussion} discusses the relationship between the measured $f_{\text{esc}}$ and potential indirect indicators of LyC escape. Sec.~\ref{sec:conclusions} summarizes our key conclusions. Throughout this work, we assume a flat $\Lambda$CDM cosmology with $H_0$ = 67.7 km s$^{-1}$ Mpc$^{-1}$ and $\Omega_m$ = 0.307 \citep{Planck2020} and the \cite{Chabrier2003} initial mass function. All magnitudes are expressed in the AB system \citep{Oke1983}.

\section{Observations and sample selection}\label{sec:data} 

The Hubble Frontier Field cluster Abell 2744 (hereafter A2744) is an ideal field for studying LyC emission, partly due to the lensing magnification of the cluster, which averages a factor of 3.5 at $z = 3$ \citep{Prieto-Lyon2023, Bergamini2023}. This magnification enables the detection of fainter galaxies \citep[$M_{\text{UV}} \approx -18$,][]{Smail1997}, which are considered analogs of cosmic reionizers.

The field was observed by the HST program 13389 (PI B. Siana), which provided WFC3/UVIS filter observations in the F336W and F275W bands \citep[8 orbits each,][]{Alavi2016}. These filters cover LyC emission for galaxies at $3.06 \lesssim z \lesssim 4$ and $2.4 \lesssim z \lesssim 3.05$, respectively.

Deep JWST near-IR spectra were obtained through the GLASS-JWST-ERS program (JWST-ERS-1324: PI T. Treu) using NIRISS and NIRSpec high-resolution observations \citep[e.g.,][]{Boyett2022, Prieto-Lyon2023, Mascia2024_GLASSrel,Watson2025}, UNCOVER program \citep{Bezanson_2024,Price2025}, as well as DDT NIRSpec prism observations (PID 2756, PI: W. Chen).

\begin{figure*}[ht!]
\centering
\includegraphics[width=0.8\textwidth]{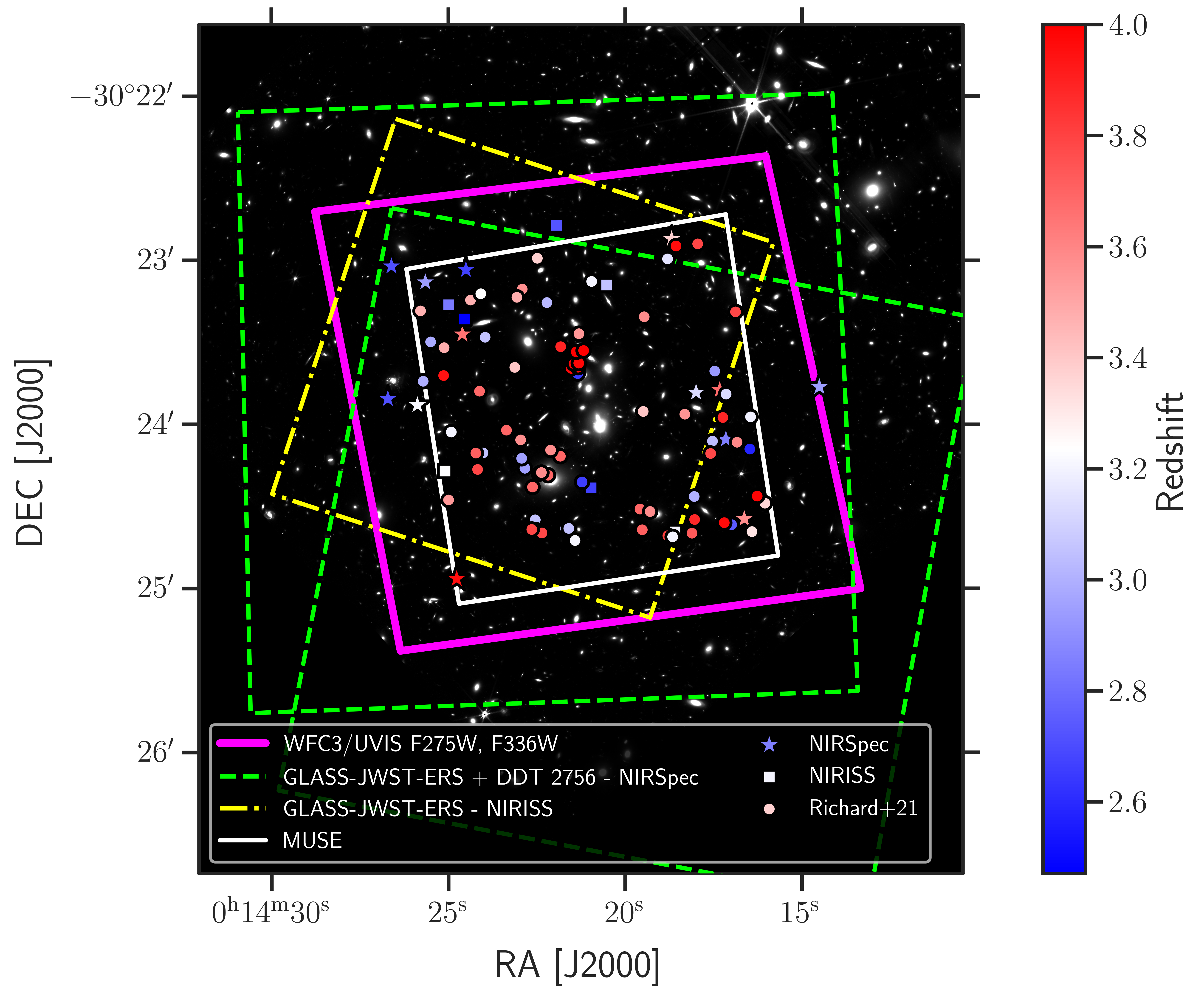}
\caption{LyC emitter candidates in the A2744 field, detected with NIRSpec (stars), NIRISS (squares), and MUSE (circles). Blue symbols represent galaxies detectable in WFC3/UVIS F275W at LyC wavelengths, while red indicates galaxies where the LyC is redshifted to the WFC3/UVIS F336W bandpass. We are also showing the footprints of HST WFC3/UVIS F275W and F336W pointings, GLASS-JWST-ERS and DDT 2756 observations, along with VLT-MUSE observations \citep{richard2021} in the A2744 cluster field. They are superimposed on the JWST NIRCam F150W image from the UNCOVER program. 
\label{fig:field}}
\end{figure*}

Additionally, VLT-MUSE observations of the region provide UV rest-frame diagnostics at $z \sim 3$, including data on the presence, strength, and shape of \lya\ emission \citep[][]{richard2021, Bergamini2023}. In Fig.~\ref{fig:field}, we present the footprints of the GLASS-JWST-ERS and DDT programs, along with the MUSE and WFC3/UVIS filter pointings.

In the following sections, we will provide detailed descriptions of the available observational dataset.

\subsection{HST and JWST imaging}\label{sec:data_hst_jwst}

We used images obtained with HST WFC3/UVIS F275W (2435-3032 \AA) and F336W (3096-3639 \AA) from the HST program PID13389 (PI: B.Siana), centered at RA=3.5832679, DEC=-30.3880082. The observations were carried out over a total of 8 orbits per filter \citep{Alavi2016}. 
%This pointing covers a highly clustered region of the field, resulting in magnified sources. 
The Lyman limit for galaxies at $z > 2.4$ falls just above the red end of the F275W filter, and for sources at $z > 3.06$, it falls within the F336W filter. Thus, these two filters are ideally suited for observing LyC emission without contamination from non-ionizing photons for galaxies in the ranges $2.4 \leq z < 3.06$ and $3.06 \leq z \leq 4$, respectively. This setup allows us to observe the LyC emission up to $\sim29-30$ AB magnitude effectively.

In addition, this field also has multiple bands of observations from the HST archive, including HST F435W, F606W, F775W and F814W filters \citep{Merlin2016}, as well as new JWST/NIRCam imaging in filters F115W, F150W, F200W, F277W, F356W, F410M, and F444W from the UNCOVER (GO 2561; PI I. Labbé) program \citep{Bezanson_2024}. 
These datasets enable us to infer accurate physical properties of the galaxies (e.g., stellar mass, SFR, $\beta$ slope), and to derive quantities such as $f_{\text{esc}}$. The analyzes of this work are mainly based on these datasets. Specially, we will use F606W as the non-ionizing band for comparison to the ionizing flux.
%In addition, this field also has multiple bands observations from HST archive and new JWST observations obtained by the GLASS-JWST ERS, UNCOVER and JWST-DDT programs \citep{Treu2022,bezanson2022,Paris2023}, which can be used to characterise the SED of the galaxies and analyzing the non-ionizing properties of the sample, e.g. $\beta$ slope, $M_{1500}$ and $f_{\text{esc}}$. 

\subsection{MUSE spectroscopic observations}\label{sec:MUSE_data}

\cite{richard2021} presented integral-field spectroscopic observations of 12 massive clusters, conducted using the Multi Unit Spectroscopic Explorer (MUSE). Observations of the A2744 cluster were carried out as part of the MUSE Guaranteed Time Observations (GTO) survey under PIDs 094.A-0115, 095.A-0181, and 096.A-0496. These observations cover the wavelength range from 4750 to 9350 \AA (as rest-frame UV for $z\sim3$), with a resolution varying from 2000 to 4000. The total exposure time for the A2744 cluster is 20 hours. The released data include a comprehensive redshift catalog of galaxies, confirmed by either emission lines, such as \lya, or continuum features, spanning a wide redshift range from $z = 0$ to $z = 6$. Quality flags are also provided to assess the reliability of the redshift determinations.

\subsection{JWST spectroscopic observations}\label{sec:JWST_spec}

%Due to the outstanding performance of JWST spectroscopic observations, the sources selected from JWST have high reliability. 
The GLASS-JWST ERS program provided high-resolution NIRSpec spectra in grating mode with a resolution of approximately 2000-3000, covering a wavelength range from 1 to 5.14 $\mu$m. The total exposure time was 4.9 hours. Additionally, the JWST-DDT program (PID 2756, PI: W. Chen) obtained continuous wavelength coverage from 0.6 to 5.3 $\mu$m using the prism configuration, which has a resolution of approximately 30-300 \citep{jakobsen2022, Treu2022, Mascia2024_GLASSrel}. The exposure time for the JWST-DDT observation was 1.23 hours. Together, these two programs yielded a total of approximately 200 unique sources with secure spectroscopic redshifts ranging from 0 to 10. Detailed information on source selection, data reduction, and analysis can be found in \cite{Treu2022, Mascia2024_GLASSrel}. 

As part of the GLASS-JWST ERS program, additional observations were carried out using the JWST/NIRISS WFSS mode, which enables the detection of strong emission-line galaxies within the field of view without the need for preselected targets. These observations provide low-resolution (R $\sim$ 150) spectra covering wavelengths from 1.0 to 2.2 $\mu$m, obtained using the F115W, F150W, and F200W blocking filters with a total integration time of ~18 hours. The data reduction process is described in \cite{Treu2022}.
\cite{Boyett2022} presented an initial redshift catalog of 76 star-forming galaxies at $1<z<3.4$, where the \oiiialone\ doublet is observable within the NIRISS wavelength range. This dataset also overlaps with the MUSE observations mentioned earlier. We further incorporated archival data from the UNCOVER survey \citep{Bezanson_2024,Price2025}, expanding the sample of high-redshift sources with robust spectroscopic confirmations.

\subsection{Sample selection}\label{sec:sample_selection}

We assembled our sample by selecting sources in the redshift range of $2.4\leq z_{\rm spec} \leq 4.0$ from the combined MUSE, NIRSpec, and NIRISS spectroscopic catalogs, requiring overlap with the field of view of the HST UV observations. A total of 98 galaxies were identified. 16 galaxies were selected from the public GLASS-JWST, UNCOVER and DDT NIRSpec observations \citep{Treu2022}, with some sources analyzed by \cite{Roy2023} and \cite{Prieto-Lyon2023}; 8 galaxies in the same redshift range were identified from the NIRISS spectra \citep{Boyett_2022}; the remaining 74 galaxies were obtained from the compilation of redshifts by the VLT-MUSE instrument \citep{richard2021, Bergamini2022}. While all JWST-derived redshifts are assigned high-quality flags, we restricted the MUSE sample to galaxies with quality flags 2 or 3, corresponding to `solid' or `secure' sources.

\begin{figure}[ht!]
\centering
\includegraphics[width=0.9\linewidth]{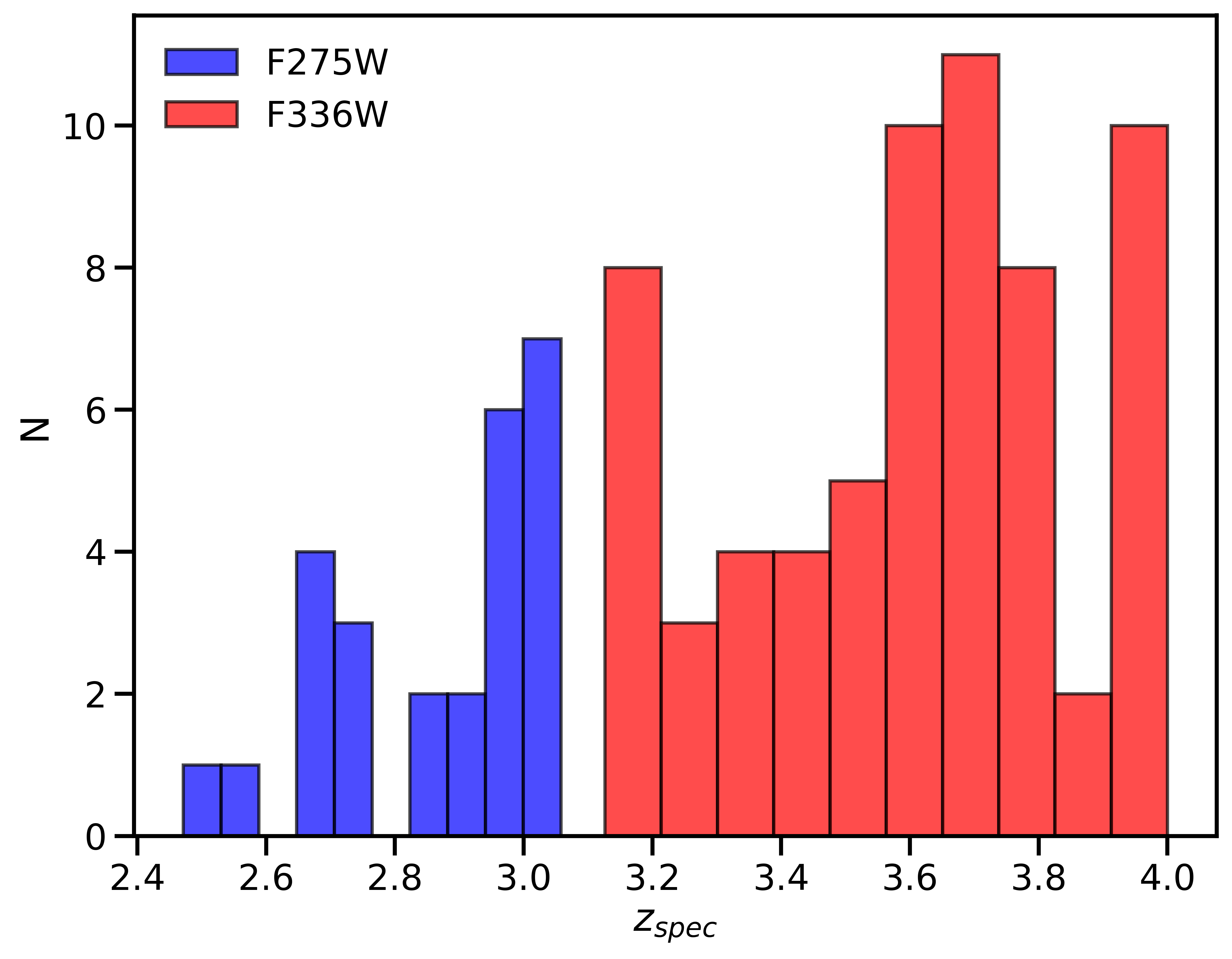}
 \caption{Redshift distribution of the 91 spectroscopically confirmed sources in the A2744 field for which we searched for LyC emission. The blue distribution represents the sources where LyC is located in the UVIS/F275W filter, while the red distribution represents those potentially detectable in the F336W filter.}
 \label{fig:redshift}
\end{figure}

At last, our final sample consists of 91 sources with secure spectroscopically confirmed redshift. Of these, 26 lie within the $2.4\leq z_{\rm spec} < 3.06$ range (corresponding to LyC flux in the WFC3/UVIS F275W filter), while 65 fall within the $3.06 \leq z_{\rm spec} \leq 4.0$ range (corresponding to LyC flux in the WFC3/UVIS F336W filter). The redshift distribution is shown in Fig.~\ref{fig:redshift}.

\section{Method}\label{sec:method}

\subsection{NIRISS spectral re-extraction}\label{sec:niriss}

To enable the measurement of rest-frame optical emission lines, we re-extracted the NIRISS grism spectra for all galaxies in our spectroscopic sample. Due to the different footprints of MUSE and NIRISS (see Fig.~\ref{fig:field}), this was only possible for the 71 sources (out of a total of 91) falling within the NIRISS footprint.

The JWST-NIRISS spectra were re-extracted following a similar method to that detailed in \cite{Roberts-Borsani2022b} and \cite{Boyett_2022}, with the differences summarized below.
We used the latest reference files (``jwst\_1123.pmap'', which includes in-flight calibrations), and version 1.11 of the Grism Redshift \& Line software \citep[\textsc{grizli},][]{Brammer2019} to reduce and process the raw data.
A flux-weighted stack was constructed from the direct NIRISS images, at a 30\,mas pixel scale, and used to generate a source segmentation map.
For each detected source, \textsc{grizli} was used to model the dispersion of a simple flat continuum in each of the three filters and two position angles, thereby generating a contamination model for the full $2\farcm1\times2\farcm1$ field of view.
Given the uncertainties associated with modeling the contamination of nearby sources, we assumed a factor of 0.2 for down-weighting contaminated pixels \citep[see][for further details]{Roberts-Borsani2022b}.

To extract the galaxies, we modeled all 2D grism exposures for an individual source simultaneously, avoiding any additional uncertainties from stacking, such as cross-dispersion offsets, whilst maintaining the same signal-to-noise ratio (S/N).
We created a set of templates using the Flexible Stellar Population Synthesis (FSPS) models of \cite{Conroy2018}, including a dedicated set of individual emission line templates, allowing for varying line ratios.
We also used the ancillary data provided by the dedicated imaging in this field to normalize the dispersed flux in each of the NIRISS filters, and to constrain the overall shape of the continuum.
For each galaxy, we therefore derived the best-fitting linear combination of templates that match the observed data when dispersed onto the sky, through non-negative least squares.
Given the low spectral resolution of JWST-NIRISS ($R\approx150$), we allowed the redshift to vary by only 1\% relative to the original $z_{{\rm spec}}$ from the catalogs (see previous section), to account for small errors in the wavelength calibration.
% opt not to fit for the redshift of the source, and instead fix the redshift of the templates to the previously-derived $z_{{spec}}$.
The 1D spectrum of each source was then optimally extracted following \cite{Horne1986}, using the position and morphology of the source in the direct image as a reference.

\subsection{HST/UVIS photometry}\label{sec:HST_phot}

We performed photometric measurements on the F275W and F336W images for sources at redshifts lower and higher than 3.06, respectively, centering on the positions retrieved from the redshift catalog for each object. The zero-points were obtained from the HST WFC3/UVIS photometric documentation\footnote{\url{https://hst-docs.stsci.edu/wfc3dhb/chapter-9-wfc3-data-analysis/9-1-photometry}}. We initially chose an aperture with a radius of $0\farcs18$, so that the diameter is approximately twice the PSF FWHM. Using \texttt{photutils}, we selected sources with a S/N greater than 2 in our sample. To ensure a reliable S/N, we masked the central region and randomly placed apertures of the same size in a $2\farcs \times 2\farcs$ region around each target position to analyze the noise level. 

Subsequently, we visually inspected the thumbnails of these sources for any possible small offsets in the LyC images. Given our focus on ($f_{\text{esc}}$), which is a key parameter for assessing LyC leakage in early galaxies, it was also important to measure the non-ionizing flux for comparison. We applied a larger aperture size ($r=0\farcs30$) to measure the global flux in the F606W band for the LyC candidates, allowing us to estimate the global escape fraction when the exact location of the LyC-leaking path is unclear \citep{grazian2016}. No aperture correction was applied when the apertures adequately fit the specific targets. The final photometric results are listed in Table \ref{tab:res_2lyc}. 

\subsection{UV stacking }\label{sec:stack}

Since most of our objects showed no detection in the band probing the LyC, we performed stacks to impose more stringent constraints on the upper limit of the LyC escape fraction. The stacking procedure averages and smooths out the LyC emission from all individuals in each subgroup.%and smooths out variations in the IGM transmission along different lines of sight. 
To do this, we created $3\farcs \times 3\farcs$ image stamps for each object in either the F275W or F336W bands, depending on the redshift of the source. The central positions were determined by the centers of the spectroscopic observations.

We separate the sources into several subgroups based on different criteria. Initially, we divided the sample into two groups simply according to their redshifts and availability of appropriate filter coverage for LyC detection. The objects with $2.4<z\leq3.06$ belong to the lower redshift group, where LyC falls within the F275W filter, while the higher redshift group ($3.06<z\leq4.0$) would have LyC detected the F336W filter. Additionally, we subdivided the sources based on other properties, such as the strength of \lya\ emission and the O32 ratio. The details of these additional criteria will be discussed in Sec.~\ref{sec:discussion}. 

To create the stacks, we proceeded as follows: we summed the image stamps for all objects in each subgroups, then normalized the flux of the stacks to the individual image flux scale. To reduce the impact of outliers, we applied a 3-sigma clipping to exclude the brightest and faintest pixels during the stacking process. We also tested the stacking process without clipping and found that the results did not change significantly, indicating that we did not miss any potential detections. These steps were performed for the F275W, F336W, and F606W bands.
%To obtain the stacks we then proceeded as follows: we summed the stamps of all the objects in each subgroup by each band, and created the stacked images by averaging the flux to correspond the flux of individual image. To reduce the effect of outliers, we applied a 3sigma-clipping to exclude the brightest and faintest pixels during the stacking. We processed the same steps in the F275W, F336W and F606W band.

We then measured the flux, analyzed the noise levels (see Sec.~\ref{sec:HST_phot}), and calculated the value at the 2-sigma dispersion level as the upper limit of the stacked images. The results are discussed in Sec.~\ref{sec:results_stack}. 

\subsection{\texorpdfstring{The escape fraction ($f_{\text{esc}}$) measurement}{The escape fraction measurement}}\label{sec:fesc_meas}

To estimate the escape fraction, $f_{\text{esc}}$, we calculate the ratio of the flux of non-ionizing photons at wavelengths $\lambda < 912$ \AA\ to the flux at $\lambda = 1500$ \AA\ rest-frame. The escape fraction is defined as:

\begin{equation}
\label{equ:abs}
f_{\text{esc}}=\frac{f_{\rm LyC,obs}}{f_{\rm UV,obs}}\frac{L_{\rm UV,intr}}{L_{\rm LyC,intr}}10^{-0.4A_{\rm UV}}e^{\tau_{\rm IGM}},
\end{equation}
\noindent
where $f_{\rm LyC,obs}/f_{\rm UV,obs}$ is the observed flux ratio between ionizing and non-ionizing radiation, and $L_{\rm UV,intr}/L_{\rm LyC,intr}$ is the intrinsic luminosity ratio. The parameter $A_{\rm UV}$ represents the dust attenuation in the non-ionizing continuum, calculated using the \cite{Calzetti2000} law, while $\tau_{\rm IGM}$ is the IGM transmission, typically evaluated around 800-900 \AA.

In this paper, we employ two methods to estimate $f_{\text{esc}}$. The first method assumes fixed values for the intrinsic luminosity ratio and dust attenuation across a sample of galaxies, while the observed flux ratio is directly calculated from photometric measurements in the appropriate filters for LyC and the non-ionizing continuum. The intrinsic luminosity ratio, $L_{\rm UV,intr}/L_{\rm LyC,intr}$, is derived from standard stellar population models \citep{Bruzual2003, Grazian2017, Rivera-Thorsen2022}. Depending on different assumptions about burst ages, initial mass functions (IMF), and metallicities, this ratio generally ranges from 3 to 7 \citep[e.g.,][]{Siana2007, Siana2010, Naidu2018, Rivera-Thorsen2022}. In this first approach, we apply a fixed intrinsic ratio of $L_{\rm UV,intr}/L_{\rm LyC,intr} = 3$ and a dust attenuation value of $A_{\rm UV} = 0.2$, which are often considered valid for young, low-mass galaxies \citep{steidel2001,grazian2016,Marchi2018}.

%For the galaxies at $z\sim2-3$, the IGM neutral hydrogen play a role in absorbing LyC photons. 
We use the model from \cite{inoue2014} to calculate the IGM transmission ($T = e^{-\tau_{\rm IGM}}$) at the wavelengths where LyC flux is detected for specific redshifts. Given the possibility of negligible CGM absorption -- due to low density or potential gaps in the CGM when LyC escape occurs -- we treat the transmission as a combined effect of both the IGM and CGM.

The second method estimates $f_{\text{esc}}$ by directly modeling the spectral energy distribution (SED) of each source, taking into account the observed LyC flux; this approach is described in detail in Sec. \ref{sec:sed_fitting}.

\section{Results}\label{sec:results}

 \subsection{Two new candidate LyC leakers }\label{sec:results_1}

 Out of 91 sources in the sample, only two have a S/N greater than 2 in either of the UV bands: MUSE4010 at $z=2.999$ in the F275W filter, and MUSE11806 at $z=3.236$ in the F336W filter. After visually inspecting the multi-band HST and JWST images and cross-referencing the redshifts with MUSE observations, we exclude the possibility that the detected LyC flux originated from lower redshift interlopers. 

Both candidates show slight offsets between the center of the LyC and non-ionizing radiation. Similar finding have been discussed in previous studies of both low and high redshifts \citep[e.g.,][]{Micheva2017, Liu2023,Komarova_2024} and suggest that a clumpy and asymmetric structures in the galaxies might facilitate LyC escape by creating channels or low-density regions \citep{Heckman2011,borthakur2014,Saldana-Lopez2022}. 
%To ensure sufficient flux is captured for estimating the global $f_{\text{esc}}$, larger apertures were used in the non-ionizing band. 

For MUSE4010, LyC is detected in the F275W filter, corresponding to a rest-frame wavelength of approximately 800 \AA\ at its redshift. Using the IGM absorption model from \cite{inoue2014}, we averaged the transmission over the 800-900 \AA\ rest-frame range and found the IGM transmission to be $T_{IGM} = 0.36\pm 0.3$. Under the fixed assumption model, we estimate the escape fraction to be $f_{\text{esc}} = 0.90^{+0.07}_{-0.86}$, indicating a significant case of LyC leakage, similar to those reported in previous studies at $z\sim3$ \citep[e.g.,][]{Vanzella2012, Saxena2022}. Errors are propagated through standard linear propagation assuming uncorrelated variables. Detecting LyC flux at high redshift may indicate low-density conditions or voids in the ISM that facilitate LyC escape. However, the IGM conditions at $z\sim3$ are highly stochastic and vary depending on the line of sight. In the case of MUSE4010, it is possible that the galaxy encountered an exceptionally transparent pathway, thus significantly reducing the impact of IGM absorption.
 
MUSE11806 shows a more compact morphology. Its IGM transmission was calculated using the 880-910 \AA\ rest-frame range, resulting in a transmission value of $T_{IGM} =0.45\pm0.33$. This wavelength range selection accounts for the fact that the F336W filter used for MUSE11806 is located closer to the Lyman limit than the F275W filter employed for MUSE4010. The estimated escape fraction for this galaxy is $f_{\text{esc}} = 0.6^{+0.37}_{-0.56}$. It is worth noting that the intrinsic luminosity ratio can vary depending on the stellar population model used, typically falling below 7 for high-redshift star-forming galaxies \citep{vanzella2010b}. For example, if we assume an intrinsic luminosity ratio of $L_{1500}/L_{900} = 5$, the escape fraction increases to $f_{\text{esc}} \sim 1$.

The results of our $f_{\text{esc}}$ measurements are summarized in Table~\ref{tab:res_2lyc}. The uncertainties in $f_{\text{esc}}$ account for both photometric errors and variations in IGM absorption, while other parameters in Eq. \eqref{equ:abs} are adopted from models without additional error contributions. %One of our candidates shows a very high escape fraction ($f_{\text{esc}} = 0.9 \pm 0.8$), while the other has $f_{\text{esc}} = 0.22 \pm 0.15$. 
%Previous studies have demonstrated that $f_{\text{esc}}$ for intermediate-redshift individual LyC leakers can vary significantly. For example, Ion1 ($z=3.794$) has $f_{\text{esc,rel}} \sim 0.3$ \citep{Ji2020}, whereas Ion3 ($z=4.0$) is estimated to have $f_{\text{esc}} \sim 0.6$ \citep{Vanzella2018}. In \citet{Saxena2022}, $f_{\text{esc}}$ ranges from 0.14 to 0.85 for galaxies at $3.11 < z < 3.53$. Furthermore, \cite{Liu2023} reports five individual LyC leakers with escape fractions between 0.4 and 0.8. It is important to note that $f_{\text{esc}}$ estimates are influenced by model assumptions, and the wide range of values can also result from sample selection and incompleteness.

\begin{figure}[ht!]

\centering
\includegraphics[width=\linewidth]{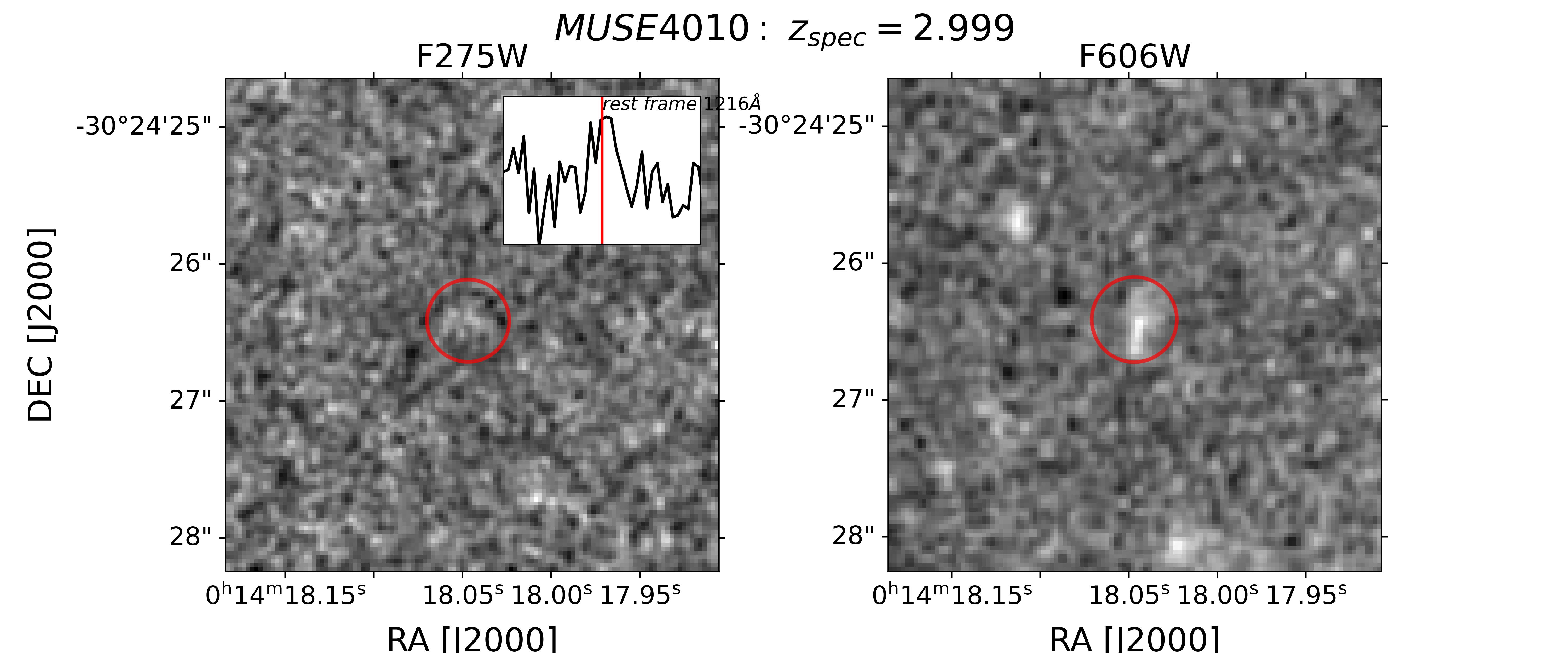}
\includegraphics[width=\linewidth]{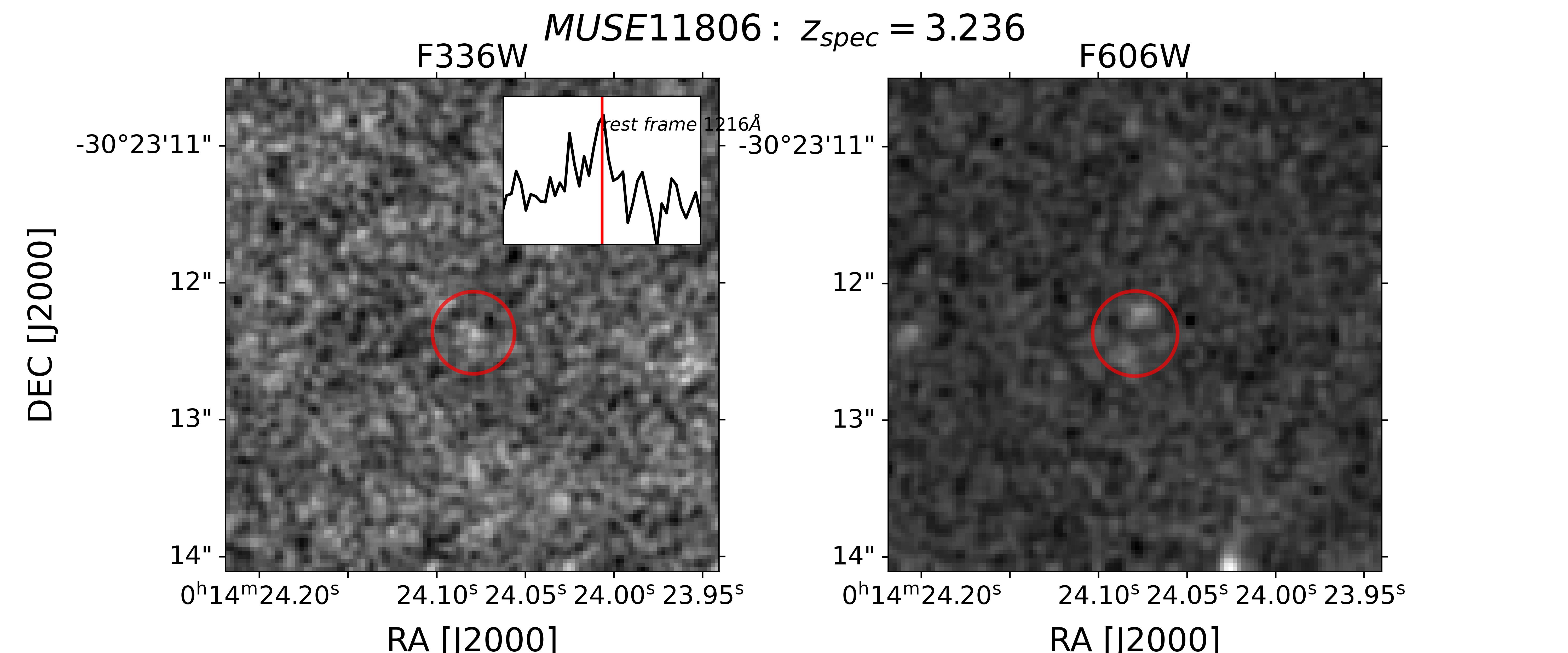}
\caption{Smoothed ionizing images (F275W or F336W) and non-ionizing images (F606W) of the two LyC candidates. The red circles indicate the apertures used for photometry, with radii of $0\farcs18$ for F275W, $0\farcs09$ for F336W, and $0\farcs30$ for F606W. The insets in the left panel present the \lya\ emission profile of each corresponding candidate. The \lya\ spectral quality flags are 2 for MUSE4010 and 3 for MUSE11806.}\label{fig:2LyC}
\end{figure}

%We put 300 random apertures with the same size on the stamps centered by each candidate. The flux distribution show that both our individual candidates are detected with S/N>2 (Fig.~\ref{fig:f_dis}). 

%\begin{figure*}[ht!]
%\includegraphics[width=0.5\textwidth]%{images/flux_distribution_random_aper_with_bkgsub_275.jpg}
%\includegraphics[width=0.5\textwidth]%{images/flux_distribution_random_aper_with_bkgsub_336.jpg}
%\caption{Left: MUSE4010: The histgram is the distribution of flux in the 300 apertures. The solid red vertical line is the detection. the gray vertical lines represent 2$\sigma$ and 3$\sigma$, while the red dashed line is constrained by a few specific apertures which have S/N>3 detection (This is not solid). Right: The same but for MUSE11806.
%\label{fig:f_dis}}
%\end{figure*}

%The photometric results are listed in Table 1.

\begin{table*}[]
\footnotesize
 \centering
 \begin{tabular}{ccccccccccccc}
 \hline\hline
 ID & RA & DEC &$z_{spec}$ &LyC     & F606W     & E(B-V)& $M_{1500}$  & $\beta$ & $M_{*}$   & $f_{\text{esc}}$&$f_{\rm esc,SED}$\\
   &[deg]&[deg]&      &$r=0\farcs18$& $r=0\farcs30$ &    &        &     &[$\textrm{log}\ M_{\odot}$]&     &\\
 \hline
 4010& 3.57519 &-30.40735&2.998 &$28.38^{+0.21}_{-0.21}$ & 28.4& $0.24^{+0.04}_{-0.04}$ &$-18.1^{+0.1}_{-0.1}$&$-2.42^{+0.05}_{-0.05}$ & $7.73^{+0.10}_{-0.10}$ & $0.9^{+0.07}_{-0.86}$ & $0.88^{+0.07}_{-0.07}$ \\
 11806 & 3.60033 &-30.38675 & 3.236&$29.60^{+0.5}_{-0.5}$ & 28.12&$0.15^{+0.02}_{-0.02}$& $-17.81^{+0.11}_{-0.11}$& $-1.78^{+0.19}_{-0.19}$ & $7.07^{+0.05}_{-0.05}$& $0.6^{+0.37}_{-0.56}$& $0.59^{+0.02}_{-0.02}$\\
 \hline
 \end{tabular}
 \caption{Photometric results and the physical properties parameters estimated by the best-fit SED model of the two candidates. The S/N of two sources are 2.07 and 2.36. $f_{\text{esc}}$ is calculated by the intrinsic model while $f_{\rm esc,SED}$ is derived from the SED fitting. $M_{1500}$, $\beta$ and $M_{*}$ have been corrected by lensing effect.
 \label{tab:res_2lyc}}
\end{table*}

\begin{table*}[ht]
\footnotesize
 \centering
 \begin{tabular}{ c c c c c}
 \hline\hline
 Group & number &LyC & F606W & $f_{\text{esc}}$\\
   & &$r=0\farcs30$ & $r=0\farcs30$ & \\
 \hline
 F275W stack & 26& >29.86 &26.52&<0.06 \\
 F336W stack & 65& >31.22 &27.56&<0.05 \\
 LAE stack & 13& >30.59 &28.25&<0.15 \\
 O32 stack & 14& >29.94 &26.52&<0.06\\
 \hline
 \end{tabular}
 \caption{Photometric results and $f_{\text{esc}}$ upper limits of the stacks. The second column represents the number of galaxies we stacked in the process. The 2$\sigma$ upper limit magnitudes for the stack of different subgroups are constraint by aperture with $r=0\farcs3$ to avoid the omission of flux due to the potential offsets. The $f_{\text{esc}}$ of the stacks are under the assumption of an average IGM transmission $T_{\rm IGM} = 0.33$, dust attenuation E(B-V)=0.2, and $L_{\rm 1500,intr}/L_{\rm 900,intr} = 3$.}
 \label{tab:phot_stack}
\end{table*}

\subsection{LyC $f_{\rm esc}$ and physical properties from SED fitting}\label{sec:sed_fitting}

To determine the physical properties of LyC candidate leakers and obtain a second estimate of their $f_{\text{esc}}$, we employed SED modeling. We combined photometric data from various optical and near-infrared bands, including HST filters F435W, F444W, F606W, F775W, and F814W, along with JWST filters F105W, F115W, F150W, F200W, F277W, F356W, and F410W.

\begin{figure}[ht!]
\includegraphics[width=\linewidth]{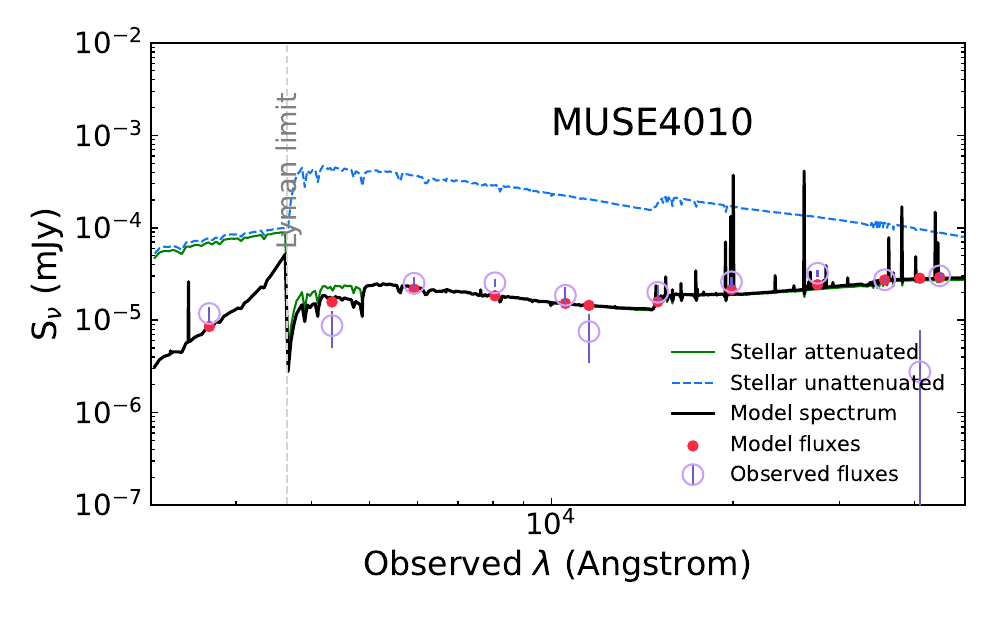}
\includegraphics[width=\linewidth]{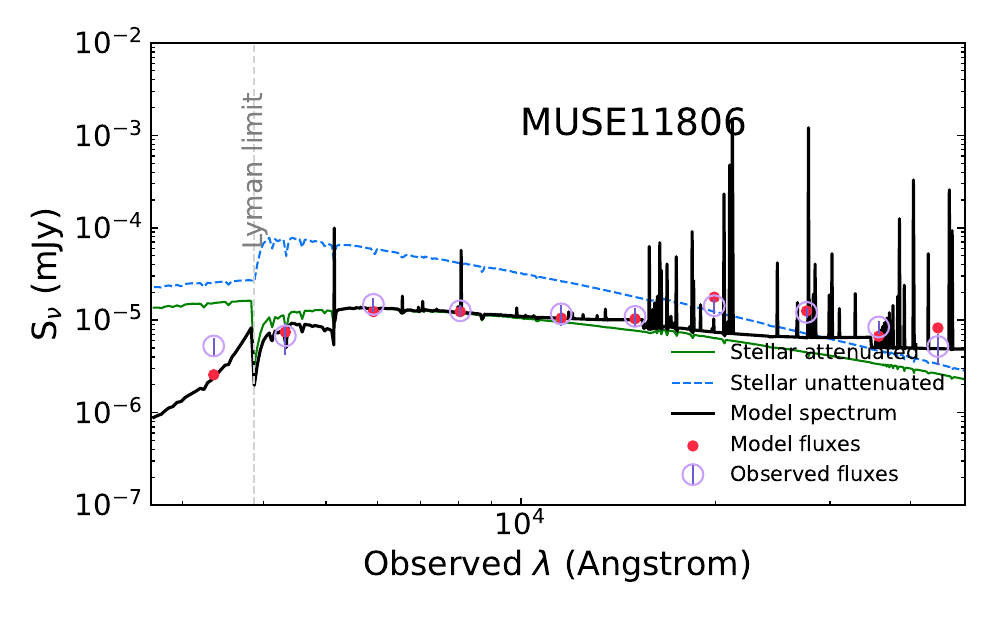}
\caption{The SED fitting results of the two objects. The purple circles are the observed flux, while the red dots are the model fluxes. The black curve is the best-fit spectrum. The unattenuated and attenuated stellar spectrum are also exhibited by green solid line and blue dashed line, respectively.
\label{fig:sedfit}}
\end{figure}

For the SED fitting, we utilized the Code Investigating GALaxy Emission \citep[CIGALE;][]{Boquien2019,Burgarella2005} with a ``double exponential'' star formation history. The dust attenuation law in CIGALE follows \cite{Calzetti2000}, incorporating a power-law attenuation curve and a UV bump to modify the slope. We used the BC03 model \citep{Bruzual2003}, with the Salpeter IMF and setting the metallicity to a low value of Z = 0.004. Additionally, we included the IGM absorption model from \cite{inoue2014}, as described in Sec.~\ref{sec:fesc_meas}. When integrating LyC observations into SED models, it is essential to consider that the distribution of neutral hydrogen and dust in the IGM and ISM is spatially complex and highly dependent on the line of sight. As a result, the sources from which we detect escaping LyC photons may reside in an over-ionized environment with higher IGM transmission (as discussed in Sec.~\ref{sec:results_1}). These conditions also increase the likelihood of reduced dust density, facilitating the passage of LyC photons. In Fig.~\ref{fig:sedfit}, we present the best-fit model spectra for the two LyC candidate leakers, while Table~\ref{tab:res_2lyc} lists their estimated physical properties. The SED analysis reveals that both LyC candidates are low-mass galaxies, with stellar masses ($M_\star$) below $10^{8} M_{\odot}$, and exhibit a relatively blue UV slope ($\beta \sim -2$). 

From the SED fitting, we derived alternative measurements of the escape fractions ($f_{\rm esc,SED}$) of $0.88 \pm 0.07$ for MUSE4010 and $0.59 \pm 0.02$ for MUSE11806, which are consistent with the $f_{\text{esc}}$ estimates derived using the general assumptions as discussed in Sec.~\ref{sec:fesc_meas}. It is important to mention that the SED fitting partially rely on the assumption of star formation history, dust extinction law and other parameters, which may not fully describe the complexity of the galaxy. It can lead to uncertainties and limitations of $f_{\text{esc}}$ estimation. For example, the average IGM model can underpredict the true transmission probability along the lines of sight with lower neutral hydrogen column densities, this effect may partially explain why the modeled LyC flux for MUSE11806 is systematically lower than the observed value.

\subsection{Constraint on LyC emission from stacks }\label{sec:results_stack}

Besides the two individual candidate LyC emitters (LCEs), the other sources in our sample do not show a significant signal (with S/N < 2) in the LyC bands. To better understand the ionizing contribution of the general galaxy population, we can establish a global estimate for $f_{\text{esc}}$ from the entire sample of undetected sources. We apply the stacking method (described in Sec.~\ref{sec:stack}) to two subgroups: the lower-redshift sources (for which the LyC emission would fall in the F275W band) and the higher-redshift sources (for which the LyC emission would fall in the F336W band).

In the lower-redshift stack, we combine data from 26 galaxies but do not find any significant flux in the stack, measuring a magnitude limit of 29.93 (S/N $\sim$ 2). For the higher-redshift subgroup, we stack 65 galaxies but still find no significant detection, allowing us to establish an even more constraining upper limit of 31.3 due to the larger sample size. The estimated $2\sigma$ upper limits of escape fractions for the two subgroups are 0.06 and 0.05, respectively. 
%The primary discrepancy between the subgroups arises mainly from their differing sample sizes.

These results indicate that most of our galaxies exhibit significantly low $f_{\text{esc}}$ (<0.1), which is consistent with previous findings. For instance, \cite{Steidel2018} reported an $f_{\text{esc}}$ of less than 0.09, derived from the composite spectra of a sample of star-forming galaxies observed through long-exposure spectroscopy with the Keck telescope. \cite{Wang2023} estimated $f_{\text{esc}}\sim 0.03$ as $2\sigma$ limit for sample of 28 galaxies in $2.5\leq z\leq 3.0$. Tight constrains on large samples were also derived by \cite{Grazian2017} who measured a stringent upper limit of $<0.02$ for the relative escape fraction from bright galaxies ($L>L_*$), while for the faint population ($L = 0.2L_*$) the limit to the escape fraction is $\lesssim$ 0.1 (see also \citealt{Boutsia2011}). 
These findings suggest that on average star-forming galaxies have low escape fractions. 

The tight constraints we obtain from stacks thus indicate that most of our galaxies have low $f_{\text{esc}}$, with the detection of just two possible individual candidate LyC emitters with high $f_{\text{esc}}$: this implies that at most only a small fraction of star-forming galaxies exhibit the conditions for a very high escape fractions. In previous works, a bimodal distribution of the $f_{\text{esc}}$ values has been suggested \citep{Nakajima2020,kreilgaard24} i.e. a (large) fraction of galaxies have a zero or very low escape fraction while a (small) fraction of galaxies have higher escape fractions. This could correspond also to line of sight effects: when the line of sight coincides with a low-density channel in the surrounding hydrogen, we see the galaxy as a high LyC leakers, while in most cases we observe galaxies though non favorable line-of sight.
In this sense the bimodal distribution could be the result of an ionisation-bounded scenario with
a small number of highly ionised, low-density channels through which the LyC photons can escape \citep{Gazagnes2020}.
Alternatively, the same bimodality could be the result of variability in the LyC production and escape that is present in many simulations  \citep{mauerhofer21,trebitsch17}.
All the above issues highlight the complexity of LyC escape mechanisms and the potential impact of this small number of star-forming galaxies on cosmic reionization.

\section{Discussion}\label{sec:discussion}

\begin{figure}[ht!]
\centering
\includegraphics[width=0.49\textwidth]{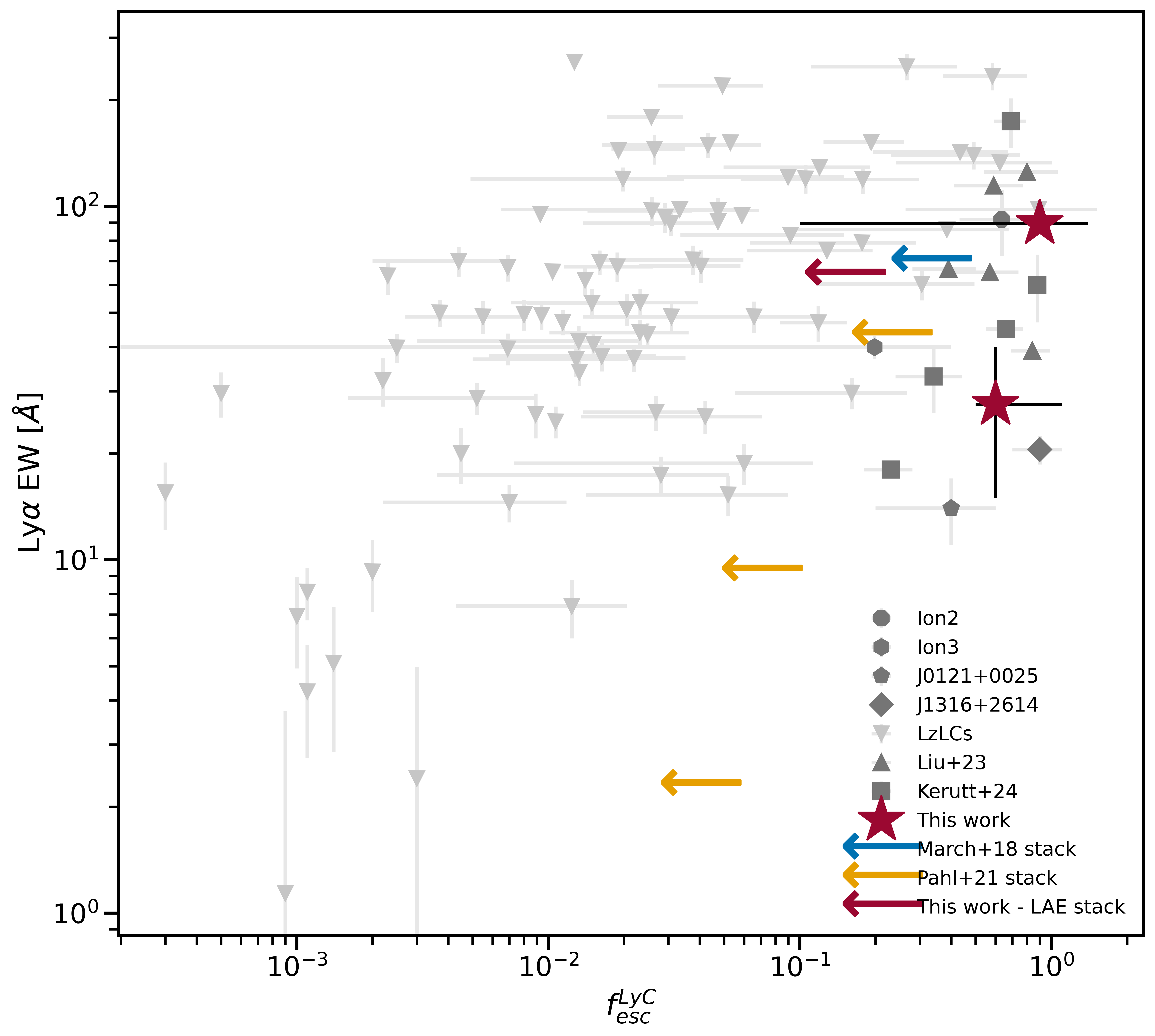}
\caption{\lya\ EW vs. $f_{\text{esc}}$. Red stars represent the two candidate leakers in our sample. Additionally, we show the $f_{\text{esc}}$ limits/values for the stacks of all the LAEs. %, the candidates in the F275W, and in the F336W.
For comparison, known leakers with \lya\ emission at $z \sim 0$ \citep{Flury2022} are shown in light gray, while those at $z > 2$ \citep{Vanzella2018, Vanzella2020, Marques-Chaves2021, Marques-Chaves2022, Liu2023, Kerutt2024} are presented in dark gray. Two limits constrained by stacks are presented in blue and yellow arrows \citep{Marchi2018,Pahl2021}.
\label{fig:Lya}}
\end{figure}

To deepen our understanding of the conditions enabling LyC emission during the EoR, it is crucial to investigate the dependencies of the LyC escape fraction on the other galaxy properties that could serve as indirect tracers. 

Studies such as those by \cite{Flury2022b} have demonstrated the effectiveness of these indirect indicators at low redshift ($z \sim 0.3$) for a galaxy sample of almost 100 sources. By examining similar correlations at $z \sim 3$, we aim to understand if the well-tested diagnostics at low redshift are still valid also at cosmic noon. We remark that the validity of the same diagnostics at cosmic noon, would endorse their suitability also in the EoR.
The primary obstacle arises from increased intergalactic medium (IGM) absorption at $z \sim 2-4$, complicating the interpretation of intermediate strength leakers and limiting detections to strong leakers ($f_{\text{esc}} \geq 0.2$). 

 Our next sections will explore the relationships between $f_{\text{esc}}$ and promising indicators such as the \lya\ emission, the O32 emission line ratio, and other physical properties of our galaxy sample.

\subsection{\texorpdfstring{\lya\ emission and O32 ratio as predictors of high $f_{\text{esc}}$}{Lya emission and O32 ratio as predictors of high fesc}}

As mentioned in the introduction, among the best indirect indicators of LyC leakage, the \lya\ recombination emission line stands out due to its high sensitivity to the environment conditions of ISM inside the galaxy along the line of sight.
%The profile of the \lya\ emission line is closely related to the processes of absorption and reemission by neutral hydrogen within the host galaxy, a phenomenon often described as a "scattering" effect \citep[e.g.,][]{Verhamme2017, Gazagnes2020}. 

Many models \citep{Verhamme2015,Verhamme2017,Dijkstra2014} predict a tight connection between the escape of \lya\ and LyC photons:
for example the characteristic \lya\ asymmetric profile (e.g. double-peak) is related to the presence of low optical depth regions in the ISM, which permit the escape of both \lya\ and LyC photons. Indeed
observations of \lya\ emissions in some LyC emitters confirmed the presence of these double-peak profiles (or even triple-peak) and demonstrated the connection with LyC leakage \citep{Vanzella2018,Rivera-Thorsen2019}. 
A broad relation between \lya\ strength and LyC emission has also been observed by several authors using stacked spectroscopy \citep[e.g.,][]{Pahl2021,Marchi2017}. Note however that recently \cite{Citro2024} challenged previous studies based on local samples, suggesting that the LyC - \lya\ relation might evolve with redshift. 

%According to many models \cite{Verhamme2015,Verhamme2017,Dijkstra2014}, strong \lya\ emission correlates to low gas column density inside the galaxies, a condition which could lead to substantial LyC leakage. Also, the characteristic \lya\ asymmetric profile (e.g. double-peak) is related to the presence of low optical depth region in the ISM, which permit the escape of both \lya\ and LyC photons. Some observations of \lya\ emissions in LyC emitters confirmed the presence of these double-peak profiles (or even triple-peak) and demonstrated the connection with LyC leakage. 

%which is consistent with the models \citep[e.g.,][]{Rivera-Thorsen2019, Izotov2018b, Vanzella2020}.
%A relation between \lya\ strength and LyC emission has %also been observed from both models\citep{Verhamme2015} and stacked spectroscopy \citep[e.g.,][]{Pahl2021,Marchi2017}. 

%Despite these promising insights, the \lya\ line's sensitivity to IGM extinction, metallicity, and stellar populations in the host galaxy can complicate its interpretation \citep[e.g.,][]{Henry2015, Yang2017a}. The neutral IGM covering increases at higher redshifts, which can significantly attenuate the \lya\ emission, particularly the blue peak \citep[e.g.,][]{Pentericci2011, Schenker2014, Yang2017b, Hayes2021}. This attenuation poses a challenge in relying solely on \lya\ emitters (LAEs) to predict the LyC $f_{\text{esc}}$, as some high-redshift Lyman break galaxies (LBGs) and LCEs exhibit minimal or no \lya\ emission \citep[e.g.,][]{Vanzella2009, Ji2020}. 

The majority of our sources (59 out of 91) including our two candidate leakers, have \lya\ in emission according to the catalog of \cite{richard2021} and 14 of them exhibit high EW (EW > 40\AA). We therefore analyze the correlation between \lya\ EW and the measured $f_{\text{esc}}$ values both for the two candidate leakers in our sample and by producing stacks of strong LAE.
%the F275W stack (27 in total, 11 sources with \lya\ emission), and the F336W stack (69 in total, 50 sources with \lya\ emission). Note that the total number of sources in the filter stacks is 14 because we are excluding the secure leakers from these. 
The two candidate leakers exhibit \lya\ EWs of $90 \pm 3$ \AA\ and $28 \pm 12$ \AA, respectively, and in particular MUSE4010 which has the highest $f_{\text{esc}}$ also has very high \lya\ EW, thus aligning with the tight link between the two quantities.
We then produced a stack of the 13 remaining strong LAEs (i.e. not considering MUSE4010), having a median \lya\ EW of 64 \AA: the stack is performed in only one band since all these LAEs are at $z\geq3.1$ and their LyC emission would fall in the F336W filter. The stack shows no signal of LyC flux and we could place a 2$\sigma$ upper limit of $f_{\text{esc}} <0.15$.
%We also note that the sources in the F275W stack presented in the previous section has shows a median \lya\ EW of $27 \pm 13$ \AA, considering the sources with \lya\ emission. Similarly, the sources in the F336W stack have an \lya\ EW of $18 \pm 11$ \AA.
\\
We compare these results with other known LyC leakers from $z = 0$ to $z = 4$ with measured \lya\ EWs.
Fig.~\ref{fig:Lya} illustrates the relationship between \lya\ EW and $f_{\text{esc}}$ for our 2 candidate leakers and our stack together with the results in the literature for individual detections \citep{Vanzella2018, Vanzella2020, Marques-Chaves2021, Flury2022, Liu2023, Kerutt2024}. 
The EW and $f_{\text{esc}}$ derived for our two candidate leakers and for the stack of strong LAEs are similar to other confirmed LyC leakers at intermediate redshift. If we consider the median $f_{\text{esc}}$ of the LzLCS+ sources within our \lya\ range (approximately between 40 and 250 \AA), we would anticipate that our sources should have an average $f_{\text{esc}}\sim 0.11$. This is slightly below our  measured limit of 0.15 meaning that we would need somewhat deeper UV data to detect the mean LyC emission. 

In general, we note that a high \lya\ EW always corresponds to a moderate or high $f_{\text{esc}}$; on the other hand objects with low  \lya\ EW can exhibit a broad range of $f_{\text{esc}}$ properties. 

%However, this property is not a stringent requirement, since the resonant nature of \lya, the individual LyC leakers are not always characterized with high EW.\\
Similarly to \lya\, the O32 emission line ratio has been extensively studied as an indicator of LyC $f_{\text{esc}}$ due to its association with density-bounded HII regions in galaxies, which can potentially facilitate LyC leakage \citep{Izotov2018b, Jaskot2013, Nakajima2014, Marchi2018,Rutkowski_2017}. For instance, Ion2 stands out as one of the strongest high-redshift known leakers with $f_{\text{esc}} > 0.5$ and O32 $ > 10$ \citep{DeBarros2016, Vanzella2016, Vanzella2020}. 
However, it seems that a high O32 ratio is necessary, but not sufficient condition for significant LyC radiation to escape the galaxy \citep{Nakajima2020}.
\cite{Naidu2018} identified a sample of strong \oiiialone\ emitters with an average $f_{\text{esc}} < 0.1$, highlighting uncertainties in using O32 as an indirect indicator. The uncertainty of O32 is primarily caused by its sensitivity to metallicity and ionization parameter \citep{Sawant2021, Bassett2019, Katz2020}.
\cite{Flury2022} propose O32 $ \geq 5$ as a threshold to identify potential LyC leakers. Note that their findings include sources with high O32 ratios that only exhibit upper limits in $f_{\text{esc}}$, suggesting once again that O32 alone may not reliably predict LyC escape due to varied escape conditions or an evolutionary sequence where LyC escape recurs at later stages when fewer early-type stars are present. 

We attempted to measure the O32 ratio for all the galaxies in our sample. Amongst the 16 sources with NIRSpec spectra available, one source lacked both \oiiialone\ and \oii\ emission lines, and two others had the lines falling outside the observed wavelength range, resulting in 13 galaxies for which O32 measurements were possible. From the NIRISS data (see also Sec.~\ref{sec:niriss}), we re-extracted spectra for 71 sources, but only 12 of them show at least one of the two emission lines with S/N > 5. This is primarily because the majority of our sources are located in the central portion of the field, where spectra are heavily contaminated by cluster galaxies (see Fig.~\ref{fig:field}). Unfortunately also our two candidate leakers are located in a region heavily contaminated by cluster galaxies, preventing us from obtaining reliable NIRISS spectra.

In total, we obtained reliable measurements (or limits) of O32 for 25 galaxies. All these galaxies exhibit detection of at least one line, typically the strongest \oiiialone\. 
%and for those lacking detection of either line, we constrained the continuum flux in the corresponding range to establish a lower limit for O32. 
We employed Gaussian profile fitting to determine the flux of each measurable emission line. The errors of line flux are propagated from the initial photons statistic and fitting process to determine O32.

\begin{figure}[ht!]
\centering
\includegraphics[width=0.49\textwidth]{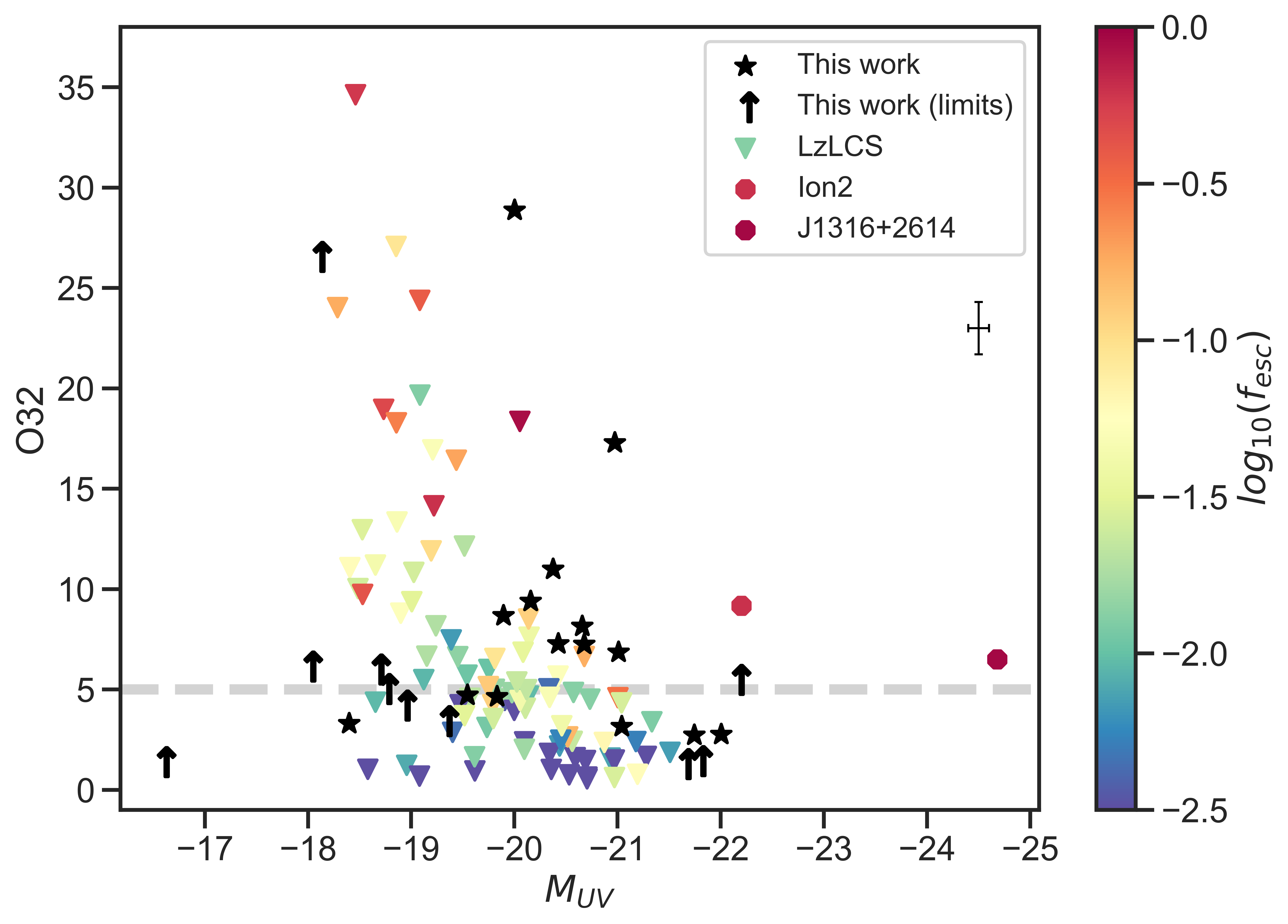}
\caption{O32 line ratio vs. $M_{\text{UV}}$. The candidates in our sample with JWST spectra showing detectable emission lines are represented by black stars if both lines are detected, or arrows if we have the \oiiialone\ measurement and \oii\ is just an upper limit. The known leakers at $z \sim 0$ and $z \sim 3$ \citep{Flury2022b, Vanzella2020, Marques-Chaves2022}, are shown color coded as a function
of their measured $f_{\text{esc}}$. The grey line (O32 = 5) indicates the threshold for LCEs proposed by \cite{Flury2022b}. A typical error is shown below the legend.
\label{fig:O32}}
\end{figure}

In Fig.~\ref{fig:O32}, we show the O32 ratio for our 25 sources, plotted vs their $M_{\text{UV}}$, corrected for magnification using the \cite{Bergamini2023} model. 
Additionally, we show individual detections of LyC leakers from the literature covering redshifts from $z = 0$ to $z = 3$, illustrating the range of O32 ratios as a function of $M_{\text{UV}}$, with colors indicating their respective $f_{\text{esc}}$ values. The O32 values of our sample align well with the distribution of previous data. 
%As mentioned in the introduction, it has been claimed that a high O32 ratio \citep[greater than 5, according to][]{Flury2022b} may be indicative of a LyC leaker with higher $f_{\text{esc}}$ altough this criterion alone is insufficient for identifying leakers. %Despite only having 8-orbit images in the LyC HST filters, some sources with O32 > 10 are intrinsically bright ($M_{\text{UV}} < -20$). For these sources, we should detect some LyC signal if it is present.

Adopting the threshold from \cite{Flury2022}, we selected the 14 sources showing O32>5 and then we stacked the UV data following the method detailed in Sec.~3.3 to detect or set a limit on their LyC emission. Since the LyC of these sources fall within either the F275W or F336W filters, we performed a two-step stacking: first computing the median flux in each filter separately, then combining the results. We obtained a non detection of the LyC flux, corresponding to an upper limit of  $f_{\text{esc}}<0.06$, assuming again $L_\nu(1500)/L_\nu(900) = 3$ \citep{steidel2001, grazian2016}, and an average E(B-V) value of 0.2.
We repeated the same exercise performed for the \lya\ EWs by considering the median $f_{\text{esc}}$ of the LzLCS+ sources within our O32 range (approximately between 5 and 20, see Fig.~\ref{fig:O32}): we predict that our sources should have an average $f_{\text{esc}}\sim 0.08$.
Thus again the expected LyC magnitude is consistent with the upper limit obtained directly from the stack of sources with high O32, implying that our UV data are currently not deep enough to probe LyC flux in such small samples.
Note that at $z\sim 3$ there are at present only two known leakers with measured O32 values \citep{Marques-Chaves2022, Vanzella2020}: these systems have similar O32 ratio as our 14 strong O32 galaxies, but much higher $f_{\text{esc}}$: one reason might be that they are both significantly brighter in terms of $M_{\rm UV}$ or in low-transparency sight lines, limiting the significance of our comparison.

%Taking into account an average IGM opacity of 0.3 as per \cite{inoue2014}, an intrinsic $L_\nu(1500)/L_\nu(900) = 3$ \citep{steidel2001, grazian2016}, and an average E(B-V) value of 0.2, the expected LyC magnitude is consistent with the upper limit obtained directly from the stack of sources with high O32 (Table \ref{tab:phot_stack}), which would suggest an $m_{LyC}$ of $\sim 31$ AB mag. 

%\textcolor{blue}{
%Even though, if we consider the LzLCS+ sources with O32 ratios comparable to ours, we would anticipate that our leakers should have on average an $f_{\text{esc}}$ lower than 0.08. The expected LyC magnitude ($m_{LyC}$) would be approximately 29.5 AB mag, which is compatible with the detection threshold of $\sim$ 29-30 AB mag achievable with the 8-orbit imaging in both the F275W and F336W filters. Additionally, at $z = 3$, the known leakers with measured O32 values \citep{Marques-Chaves2022, Vanzella2020} are significantly brighter compared to ours, limiting the extent of our comparison. Thus, while our initial estimate aligns with the threshold, the median $f_{\text{esc}}$ indicates that detecting these faint sources, despite the high O32 ratios measured, remains challenging within the current observational limits.}

\subsection{Physical properties of the candidate LyC leakers}

Investigating the physical properties of LyC emitters is crucial for understanding the mechanisms enabling the escape of ionizing radiation from galaxies and their role in the reionization of the universe. 
%Although the occurrence of reionization is well-established, the specific types of objects that controlled and dominated this process remain uncertain.
The debate over which galaxy mass regime primarily drives reionization is ongoing.
Low-mass galaxies ($M_* \sim 10^9 M_\odot$), with their weaker gravitational potentials, are more prone to removal of obscuring material by stellar winds and supernovae (SNe), thereby enhancing the escape fraction of LyC photons \citep[e.g.,][]{Razoumov2010, Wise2014, Paardekooper2015, Robertson2022}. In contrast, more massive galaxies possess the gas reservoirs necessary to sustain high star formation rate (SFR) densities. These galaxies experience less negative feedback from stellar winds and SNe, allowing for more substantial star formation and, consequently, greater LyC photon production \citep{Wyithe2013}.
\\
% Some models, based on empirical data, suggest that relatively more luminous galaxies are the main sources of LyC photons responsible for reionization \citep{Naidu2020}. Other models propose that less-luminous galaxies dominate reionization due to their steeper luminosity functions \citep[e.g.,][]{Finkelstein2019}, and even low $f_{\text{esc}}$ (average <5\%) can produce enough ionizing photons for reionization \citep{atek2024}. 
Observations of local LyC emission \citep[e.g.,][]{Izotov2018b, Flury2022, Flury2022b} tend to support the dwarf-galaxy scenario indicated by radiation hydrodynamical simulations \citep[e.g.,][]{Trebitsch2017}, as do some comprehensive models that include massive galaxies, active galactic nuclei (AGNs), and dwarf galaxies \citep[e.g.,][]{Dayal2020}.

In the LzLCS+ sample \citep{Flury2022b, Flury2022} the strongest leakers exhibit the lowest stellar masses, though still spanning a rather large range from $M_* = 10^7$ to almost $10^{10} M_\odot$; they tend to be the most compact sources, with UV half-light radii $r_e$ ranging from 0.3 to 2 kpc; finally, their $\Sigma_{\text{SFR}}$ and sSFR values suggest highly concentrated star formation regions. \cite{Chisholm2022} also identified the slope $\beta$ as one of the properties that is highly correlated with
$f_{\text{esc}}$ due to its links with dust and metallicity in the environment.
%highly effective indirect indicator. The UV slope, after adjusting for the nebular continuum, shows a strong correlation with the continuum color excess, $E(B-V)$ \citep{Saldana-Lopez2022}, indicating a likely relationship with the $f_{\text{esc}}$. 
Specifically, galaxies with $\beta = -2.11, -2.35, -2.60$ exhibit population-averaged $f_{\text{esc}} = 0.05, 0.10, 0.20$, respectively. Currently, with few and sparse detections of LCEs at $z \sim 3$ compared to the LzLCS+ sample, it is unclear if the same trends are also valid at intermediate redshift.

\begin{figure*}[ht!]
\centering
\includegraphics[width=0.49\textwidth]{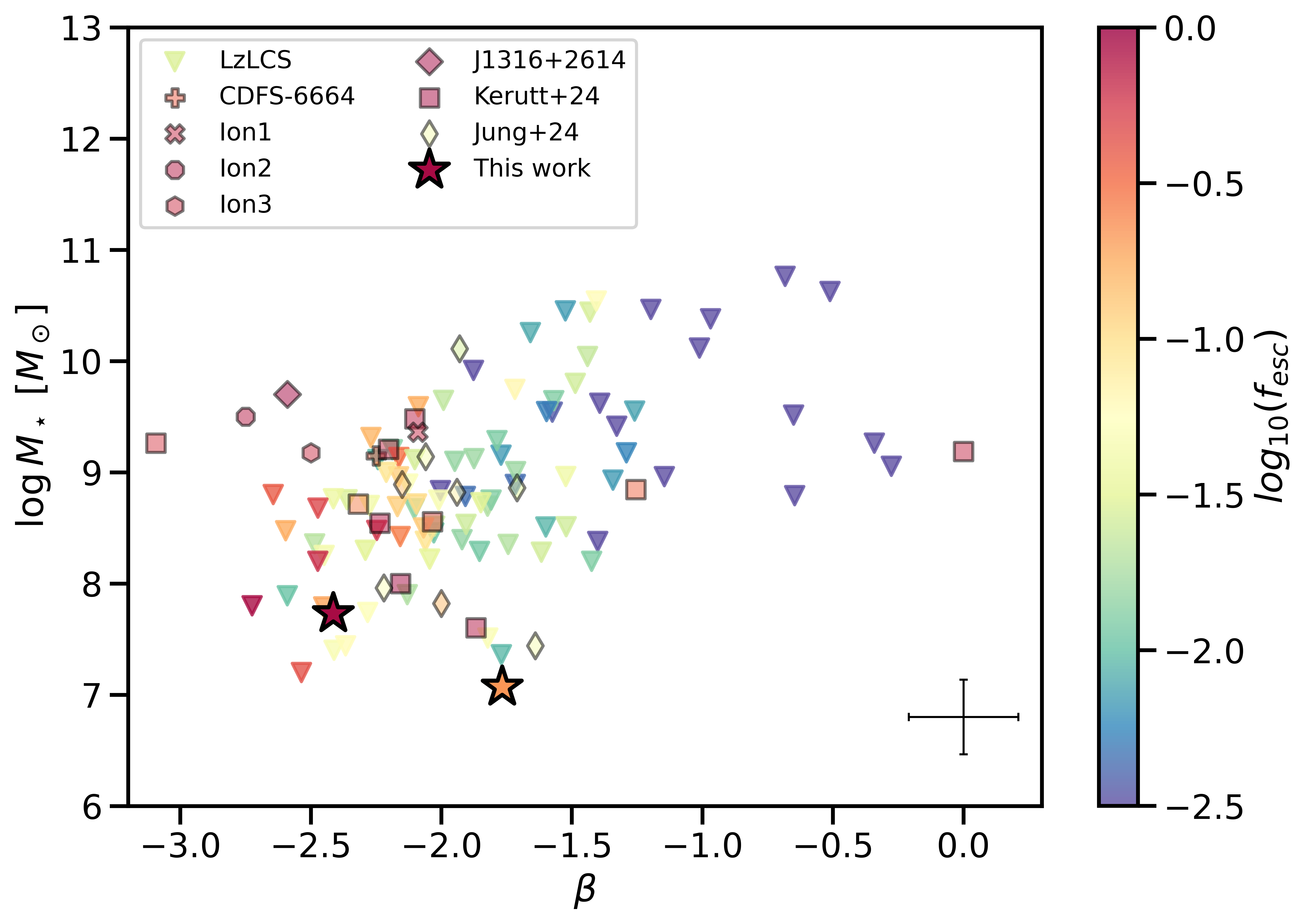}
\includegraphics[width=0.49\textwidth]{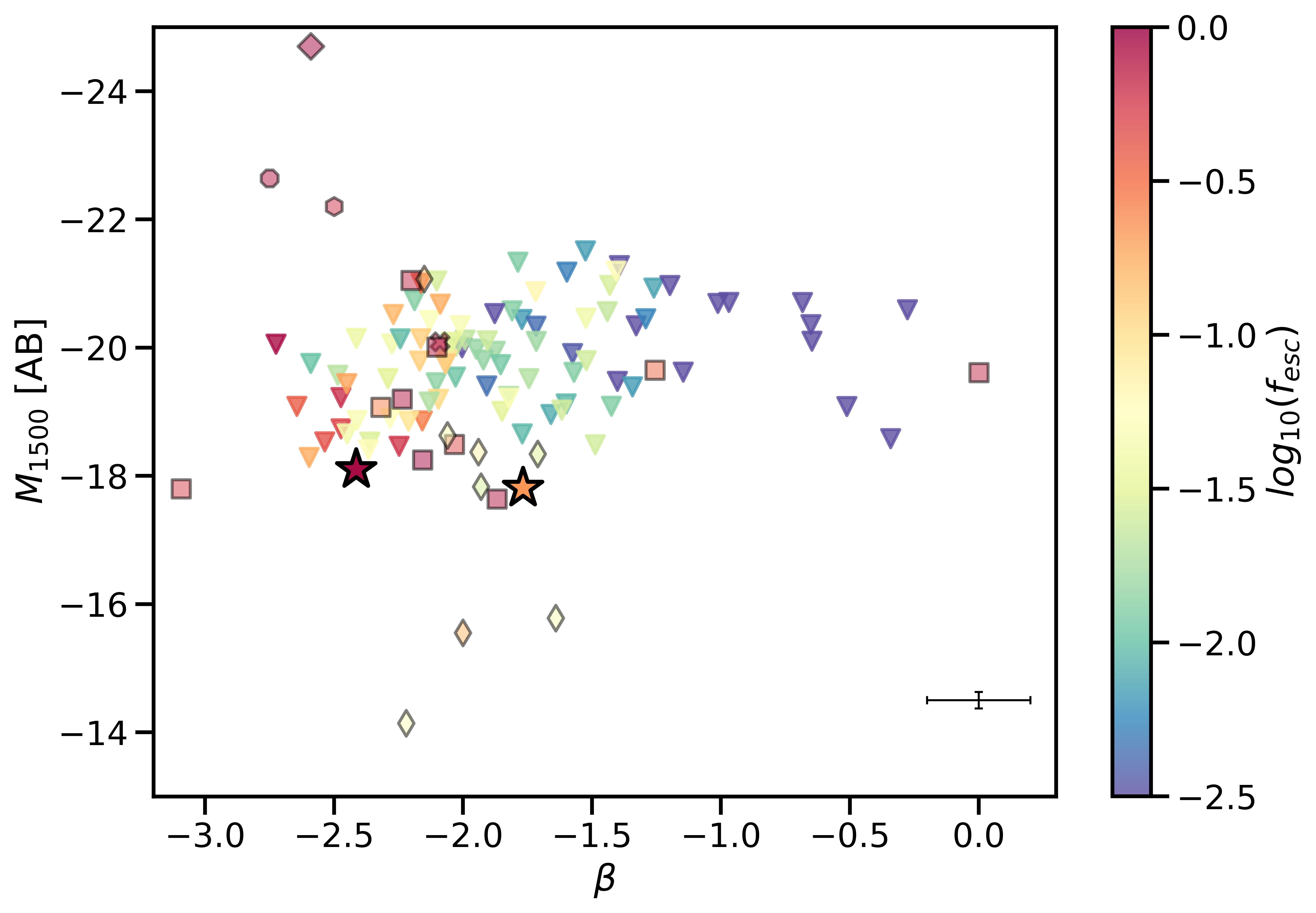}
\caption{The location of our two LyC candidate leakers (stars) on the $\beta$ vs. stellar mass (left) and $\beta$ vs. $M_{\text{UV}}$ (right) diagrams, compared to known leakers at $z \sim 0.3$ \citep[][]{Flury2022b} and $z > 2$ \citep{Vanzella2016, Vanzella2018, Vanzella2020, Marques-Chaves2022, Kerutt2024, Jung2024}. Sources are color-coded based on their $\log_{10} f_{\text{esc}}$ values. A typical errorbar is shown in the right bottom corner of each figure.\label{fig:density_plot}}
\end{figure*}

%we find that these galaxies' properties are not significantly different but rather scattered in plots such as stellar mass vs. $\beta$ slope and $M_{\text{UV}}$ vs. $\beta$ slope. 
In Fig.~\ref{fig:density_plot}, we show various physical properties of our two new z=3 candidate LyC leakers, compared to the LzLCS+ sample and to the known leakers at intermediate redshifts for which we have measured properties \citep{Vanzella2016, Vanzella2018, Vanzella2020, Marques-Chaves2022, Kerutt2024, Jung2024}. 
The two new candidates exhibit properties comparable to those of known LyC leakers with similar $f_{\text{esc}}$ at low and intermediate redshifts. Specifically, they have blue UV $\beta$ slopes, low stellar masses and faint $M_{\rm UV}$.
The intrinsic UV half-light radii of the sources, measured in the JWST/NIRCam F090W, corrected for lensing magnification \citep[$\mu$ = 2.59 and 2.23, ][]{Bergamini2023}, are 0.96 kpc and 0.61 kpc, respectively. These values confirm the compact nature of the systems, consistent with expectations for LyC emitting galaxies \citep{Flury2022b}.

Once again, we note that the scatter shown between LyC emission and physical properties (as also noted when discussing \lya\ and O32) is large and underscores the complexity and diversity of the LyC escape process across different epochs. 
%Combining observations at low and intermediate redshifts, there is a wide range in the physical properties of known leakers, without any tight correlation between each individual property and $f_{\text{esc}}$. 
For this reason, several studies have employed a multivariate approach to model the relation between
$f_{\text{esc}}$ and the physical properties of the sources, in order to effectively infer $f_{\text{esc}}$ for sources during the EoR where the direct detection of LyC flux is impossible \citep[e.g.,][]{Jaskot24a, Jaskot24b, Mascia2023, Mascia2024, Choustikov2024, Lin2024}.
Currently, such predictions are essentially calibrated using the LzLCS+ sample since, as evident from Figure ~\ref{fig:density_plot}, confirmed LyC leakers at $z=3$ are scarce, and most of them exhibit higher $f_{\text{esc}}$ than their low redshift counterparts (see also Fig.~\ref{fig:Lya}, where the offset between low- and high-redshift leakers is apparent). This is due to the higher IGM opacity at high redshift, which implies that the strong leakers are the only ones easily detectable. It could also be due to high redshift galaxies following different LyC escape physics, but substantial scatter in the data limits our ability to robustly quantify such differences or extrapolate low-z trends. Additional observations and confirmed LyC emitters at cosmic noon are thus essential to establish robust conclusions through comparison with low-redshift samples.

%\subsection{Hints of different mechanisms of LyC escape?}

%\textcolor{blue}{
%\cite{Flury2022b} discuss the variability of the LyC escape process across different galaxy types. For instance, the dominant feedback mechanism may shift from radiative feedback to mechanical SN feedback as a starburst ages or in galaxies with different masses or metallicities \citep[e.g.,][]{Jaskot2019, Kimm2019, Jecmen2023}. Additionally, the strongest LCEs may exhibit a nearly density-bounded gas geometry, while in weaker LCEs, LyC photons might escape through narrow channels \citep[e.g.,][]{Gazagnes2020, Flury2022b}. \\

%Understanding the physical properties of LyC emitters—such as stellar mass, UV half-light radii, $\Sigma_{\text{SFR}}$, sSFR, O32 ratios, $\beta$ slopes, and $M_{\text{UV}}$—is essential for comprehending the conditions that facilitate LyC photon escape. These properties differentiate LyC emitters from non-emitters and provide valuable insights into the diverse and complex processes behind LyC escape, which might infer different mechanism between low-redshift and high-redshift sources.}

\section{Summary and conclusions}\label{sec:conclusions}

In this work, we searched for new LyC candidate leakers within a sample of 91 spectroscopically confirmed star-forming galaxies in the A2744 cluster field. The sample includes galaxies with spectra obtained from JWST/NIRSpec, JWST/NIRISS WFSS, and VLT/MUSE observations. We searched for LyC flux in the HST/F275W and HST/F336W filters, corresponding to the rest-frame 800-900 \AA\ for galaxies at $2.4<z<3.06$ and $3.06 <z <4$ respectively. Additionally, we performed SED fitting using multi-band observations from HST and JWST/NIRCam to estimate the physical properties of these galaxies.

We identified two new $z\sim3$ candidate LyC leakers, with LyC flux showing a S/N > 2, and we derived their $f_{\text{esc}}$ using two approaches: direct calculation and SED model fitting. Both identified candidate leakers exhibit escape fractions greater than 0.2, with MUSE4010 showing an especially high escape fraction of approximately 0.9, consistently between the two methods. 
For the rest of the non-detections (i.e. 89 objects with S/N < 2 in the UV bands), we employed a stacking of the UV bands but found no significant signal, thereby constraining the upper limit of LyC escape for both the lower- and higher-redshift groups to be $f_{\text{esc}} \leq 0.058$. These results are consistent with previous findings in the literature from stacking analysis \citep[e.g.,][]{vanzella_2015, Rutkowski_2017, Liu2023, Kerutt2024}, which suggest that only a small fraction of star-forming galaxies have conditions leading to a conspicuous escape of ionizing photons, whereas the general population has much lower $f_{\text{esc}}$ values.
Precise $f_{\text{esc}}$ estimates remain sensitive to assumptions about stellar population, dust attenuation and IGM absorption, highlighting the uncertainties in these calculations especially at intermediate redshift.

We also explored the properties that have been proposed to be the best indirect tracers of LyC escape. Particularly we investigated the relationship between $f_{\text{esc}}$, \lya\ EW, O32 ratio, stellar mass, $M_{\text{UV}}$, and $\beta$. 
The two new candidate leakers are indeed \lya\ emitters, have blue UV slopes, are low mass, faint sources and are very compact in the UV. Unfortunately, we were unable to measure their O32 ratio from the NIRISS spectra, due to limited sensitivity. We also analyzed the spectral stacks of the strong \lya\ emitters and the sources with a high O32 ratio obtaining only non-detection, with less stringent upper limits on the $f_{\text{esc}}$, probably due to the current observational limitations of our UV data. 
%\textcolor{blue}{
%For the 14 galaxies with confirmed \lya\ emission, we find that both the EW and $f_{\text{esc}}$ are consistent with values reported in the literature, suggesting that high \lya\ EW could serve as a clue for high LyC escape. We also measured the O32 ratio for 20 galaxies with available spectra from both NIRSpec and NIRISS, obtaining reliable values or upper limits. %Fig.~\ref{fig:O32} shows that 
%Our sources align well with known LyC leakers at both high and low redshifts, particularly those with O32 > 5, indicating a possible connection between high O32 ratios and LyC escape. However, due to current observational limitations, particularly in obtaining high S/N LyC detections, definitive conclusions remain elusive.

The question of which specific types of galaxies drove reionization remains an open topic of debate. Our observations, combined with those of existing LCEs at $z\sim3$ and lower redshifts, reinforce the evidence that properties such as stellar mass, $\beta$, and $M_{\text{UV}}$ show consistent correlations with $f_{\text{esc}}$ across different redshifts. However, these relations are very scattered because LyC escape is a complex process, with multiple pathways and mechanisms enabling ionizing photon leakage at different epochs and across diverse galaxy populations \citep{Jaskot2019, Flury2022b}, and could be highly affected by the geometric distribution and line-of-sight effects. To further refine our understanding of LyC escape mechanisms and determine any possible evolution with cosmic time, deeper observations and larger samples at cosmic noon ($z=2-4$) would be needed, also probing different fields to mitigate the IGM transmission variations. We remark that it would also be extremely important to detect galaxies with more modest LyC escape fractions at this intermediate redshift, which would be essential to confirm potential correlations between LyC escape and galaxy properties.
In this way, we could construct robust predictions of LyC escape from multivariate diagnostics and apply them confidently during the epoch of reionization. 

%mention heere the upcoming UV HST data 

\begin{acknowledgements}
 We acknowledge support from the National Science Foundation of China - 12225301, INAF Large grant "Spectroscopic survey with JWST" jand from PRIN 2022 MUR project 2022CB3PJ3 - First Light And Galaxy aSsembly (FLAGS) funded by the European Union – Next Generation EU, and Postgraduate Scholarship Program under the grant of China Scholarship Council. P.W. and B.V. acknowledge support from the INAF Mini Grant `1.05.24.07.01 RSN1: Spatially Resolved Near-IR Emission of Intermediate-Redshift Jellyfish Galaxies' (PI Watson). 
 We acknowledge A. Acebron, C. Grillo, and P. Rosati for their fundamental contribution to the strong lensing analysis and results. We also extend our gratitude to the JWST and HST teams for their efforts in designing, building, and operating these transformative missions.
\end{acknowledgements}

\bibliographystyle{aa}
%\bibliography{biblio.bib} 

\begin{thebibliography}{118}
\expandafter\ifx\csname natexlab\endcsname\relax\def\natexlab#1{#1}\fi

\bibitem[{{Alavi} {et~al.}(2016){Alavi}, {Siana}, {Richard}, {Rafelski},
  {Jauzac}, {Limousin}, {Freeman}, {Scarlata}, {Robertson}, {Stark}, {Teplitz},
  \& {Desai}}]{Alavi2016}
{Alavi}, A., {Siana}, B., {Richard}, J., {et~al.} 2016, \apj, 832, 56

\bibitem[{{Bassett} {et~al.}(2019){Bassett}, {Ryan-Weber}, {Cooke}, {Diaz},
  {Nanayakkara}, {Yuan}, {Spitler}, {Me{\v{s}}tri{\'c}}, {Garel}, {Sawicki},
  {Gwyn}, \& {Golob}}]{Bassett2019}
{Bassett}, R., {Ryan-Weber}, E.~V., {Cooke}, J., {et~al.} 2019, \mnras, 483,
  5223

\bibitem[{{Bergamini} {et~al.}(2023{\natexlab{a}}){Bergamini}, {Acebron},
  {Grillo}, {Rosati}, {Caminha}, {Mercurio}, {Vanzella}, {Angora}, {Brammer},
  {Meneghetti}, \& {Nonino}}]{Bergamini2022}
{Bergamini}, P., {Acebron}, A., {Grillo}, C., {et~al.} 2023{\natexlab{a}},
  \aap, 670, A60

\bibitem[{{Bergamini} {et~al.}(2023{\natexlab{b}}){Bergamini}, {Acebron},
  {Grillo}, {Rosati}, {Caminha}, {Mercurio}, {Vanzella}, {Mason}, {Treu},
  {Angora}, {Brammer}, {Meneghetti}, {Nonino}, {Boyett}, {Brada{\v{c}}},
  {Castellano}, {Fontana}, {Morishita}, {Paris}, {Prieto-Lyon},
  {Roberts-Borsani}, {Roy}, {Santini}, {Vulcani}, {Wang}, \&
  {Yang}}]{Bergamini2023}
{Bergamini}, P., {Acebron}, A., {Grillo}, C., {et~al.} 2023{\natexlab{b}},
  \apj, 952, 84

\bibitem[{Bezanson {et~al.}(2024)Bezanson, Labbe, Whitaker, Leja, Price, Franx,
  Brammer, Marchesini, Zitrin, Wang, Weaver, Furtak, Atek, Coe, Cutler, Dayal,
  van Dokkum, Feldmann, Förster~Schreiber, Fujimoto, Geha, Glazebrook,
  de~Graaff, Greene, Juneau, Kassin, Kriek, Khullar, Maseda, Mowla, Muzzin,
  Nanayakkara, Nelson, Oesch, Pacifici, Pan, Papovich, Setton, Shapley, Smit,
  Stefanon, Taylor, \& Williams}]{Bezanson_2024}
Bezanson, R., Labbe, I., Whitaker, K.~E., {et~al.} 2024, The Astrophysical
  Journal, 974, 92

\bibitem[{{Bolan} {et~al.}(2022){Bolan}, {Lemaux}, {Mason}, {Brada{\v{c}}},
  {Treu}, {Strait}, {Pelliccia}, {Pentericci}, \& {Malkan}}]{bolan2021}
{Bolan}, P., {Lemaux}, B.~C., {Mason}, C., {et~al.} 2022, \mnras, 517, 3263

\bibitem[{{Boquien, M.} {et~al.}(2019){Boquien, M.}, {Burgarella, D.},
  {Roehlly, Y.}, {Buat, V.}, {Ciesla, L.}, {Corre, D.}, {Inoue, A. K.}, \&
  {Salas, H.}}]{Boquien2019}
{Boquien, M.}, {Burgarella, D.}, {Roehlly, Y.}, {et~al.} 2019, A\&A, 622, A103

\bibitem[{Borthakur {et~al.}(2014)Borthakur, Heckman, Leitherer, \&
  Overzier}]{borthakur2014}
Borthakur, S., Heckman, T.~M., Leitherer, C., \& Overzier, R.~A. 2014, Science,
  346, 216

\bibitem[{{Bosman} {et~al.}(2022){Bosman}, {Davies}, {Becker}, {Keating},
  {Davies}, {Zhu}, {Eilers}, {D'Odorico}, {Bian}, {Bischetti}, {Cristiani},
  {Fan}, {Farina}, {Haehnelt}, {Hennawi}, {Kulkarni}, {Mesinger}, {Meyer},
  {Onoue}, {Pallottini}, {Qin}, {Ryan-Weber}, {Schindler}, {Walter}, {Wang}, \&
  {Yang}}]{Bosman2022}
{Bosman}, S. E.~I., {Davies}, F.~B., {Becker}, G.~D., {et~al.} 2022, \mnras,
  514, 55

\bibitem[{{Boutsia} {et~al.}(2011){Boutsia}, {Grazian}, {Giallongo}, {Fontana},
  {Pentericci}, {Castellano}, {Zamorani}, {Mignoli}, {Vanzella}, {Fiore},
  {Lilly}, {Gallozzi}, {Testa}, {Paris}, \& {Santini}}]{Boutsia2011}
{Boutsia}, K., {Grazian}, A., {Giallongo}, E., {et~al.} 2011, \apj, 736, 41

\bibitem[{{Boyett} {et~al.}(2022){Boyett}, {Mascia}, {Pentericci},
  {Leethochawalit}, {Trenti}, {Brammer}, {Roberts-Borsani}, {Strait}, {Treu},
  {Bradac}, {Glazebrook}, {Acebron}, {Bergamini}, {Calabr{\`o}}, {Castellano},
  {Fontana}, {Grillo}, {Henry}, {Jones}, {Marchesini}, {Mason}, {Mercurio},
  {Morishita}, {Nanayakkara}, {Rosati}, {Scarlata}, {Vanzella}, {Vulcani},
  {Wang}, \& {Willott}}]{Boyett2022}
{Boyett}, K., {Mascia}, S., {Pentericci}, L., {et~al.} 2022, \apjl, 940, L52

\bibitem[{Boyett {et~al.}(2022)Boyett, Mascia, Pentericci, Leethochawalit,
  Trenti, Brammer, Roberts-Borsani, Strait, Treu, Bradac, Glazebrook, Acebron,
  Bergamini, Calabrò, Castellano, Fontana, Grillo, Henry, Jones, Marchesini,
  Mason, Mercurio, Morishita, Nanayakkara, Rosati, Scarlata, Vanzella, Vulcani,
  Wang, \& Willott}]{Boyett_2022}
Boyett, K., Mascia, S., Pentericci, L., {et~al.} 2022, The Astrophysical
  Journal Letters, 940, L52

\bibitem[{{Brammer}(2019)}]{Brammer2019}
{Brammer}, G. 2019, {Grizli: Grism redshift and line analysis software},
  Astrophysics Source Code Library, record ascl:1905.001

\bibitem[{{Bruzual} \& {Charlot}(2003)}]{Bruzual2003}
{Bruzual}, G. \& {Charlot}, S. 2003, \mnras, 344, 1000

\bibitem[{Burgarella {et~al.}(2005)Burgarella, Buat, \&
  Iglesias-Páramo}]{Burgarella2005}
Burgarella, D., Buat, V., \& Iglesias-Páramo, J. 2005, Monthly Notices of the
  Royal Astronomical Society, 360, 1413

\bibitem[{{Calzetti} {et~al.}(2000){Calzetti}, {Armus}, {Bohlin}, {Kinney},
  {Koornneef}, \& {Storchi-Bergmann}}]{Calzetti2000}
{Calzetti}, D., {Armus}, L., {Bohlin}, R.~C., {et~al.} 2000, \apj, 533, 682

\bibitem[{{Chabrier}(2003)}]{Chabrier2003}
{Chabrier}, G. 2003, \pasp, 115, 763

\bibitem[{{Chisholm} {et~al.}(2022){Chisholm}, {Saldana-Lopez}, {Flury},
  {Schaerer}, {Jaskot}, {Amor{\'\i}n}, {Atek}, {Finkelstein}, {Fleming},
  {Ferguson}, {Fern{\'a}ndez}, {Giavalisco}, {Hayes}, {Heckman}, {Henry}, {Ji},
  {Marques-Chaves}, {Mauerhofer}, {McCandliss}, {Oey}, {{\"O}stlin},
  {Rutkowski}, {Scarlata}, {Thuan}, {Trebitsch}, {Wang}, {Worseck}, \&
  {Xu}}]{Chisholm2022}
{Chisholm}, J., {Saldana-Lopez}, A., {Flury}, S., {et~al.} 2022, \mnras, 517,
  5104

\bibitem[{{Choustikov} {et~al.}(2024){Choustikov}, {Katz}, {Saxena}, {Cameron},
  {Devriendt}, {Slyz}, {Rosdahl}, {Blaizot}, \&
  {Michel-Dansac}}]{Choustikov2024}
{Choustikov}, N., {Katz}, H., {Saxena}, A., {et~al.} 2024, \mnras, 529, 3751

\bibitem[{{Citro} {et~al.}(2025){Citro}, {Scarlata}, {Mantha}, {Williams},
  {Rafelski}, {Revalski}, {Hayes}, {Henry}, {Rutkowski}, {Teplitz}, {Grazian},
  \& {Alavi}}]{Citro2024}
{Citro}, A., {Scarlata}, C.~M., {Mantha}, K.~B., {et~al.} 2025, \apj, 986, 184

\bibitem[{{Conroy} {et~al.}(2018){Conroy}, {Villaume}, {van Dokkum}, \&
  {Lind}}]{Conroy2018}
{Conroy}, C., {Villaume}, A., {van Dokkum}, P.~G., \& {Lind}, K. 2018, \apj,
  854, 139

\bibitem[{{Dayal} {et~al.}(2020){Dayal}, {Volonteri}, {Choudhury}, {Schneider},
  {Trebitsch}, {Gnedin}, {Atek}, {Hirschmann}, \& {Reines}}]{Dayal2020}
{Dayal}, P., {Volonteri}, M., {Choudhury}, T.~R., {et~al.} 2020, \mnras, 495,
  3065

\bibitem[{{Dayal} {et~al.}(2025){Dayal}, {Volonteri}, {Greene}, {Kokorev},
  {Goulding}, {Williams}, {Furtak}, {Zitrin}, {Atek}, {Bezanson},
  {Chemerynska}, {Feldmann}, {Glazebrook}, {Labbe}, {Nanayakkara}, {Oesch}, \&
  {Weaver}}]{dayal2024}
{Dayal}, P., {Volonteri}, M., {Greene}, J.~E., {et~al.} 2025, \aap, 697, A211

\bibitem[{{de Barros} {et~al.}(2016){de Barros}, {Vanzella}, {Amor{\'{\i}}n},
  {Castellano}, {Siana}, {Grazian}, {Suh}, {Balestra}, {Vignali}, {Verhamme},
  {Zamorani}, {Mignoli}, {Hasinger}, {Comastri}, {Pentericci},
  {P{\'e}rez-Montero}, {Fontana}, {Giavalisco}, \& {Gilli}}]{DeBarros2016}
{de Barros}, S., {Vanzella}, E., {Amor{\'{\i}}n}, R., {et~al.} 2016, \aap, 585,
  A51

\bibitem[{{Dijkstra} {et~al.}(2014){Dijkstra}, {Wyithe}, {Haiman}, {Mesinger},
  \& {Pentericci}}]{Dijkstra2014}
{Dijkstra}, M., {Wyithe}, S., {Haiman}, Z., {Mesinger}, A., \& {Pentericci}, L.
  2014, \mnras, 440, 3309

\bibitem[{{Finkelstein} {et~al.}(2019){Finkelstein}, {D'Aloisio},
  {Paardekooper}, {Ryan}, {Behroozi}, {Finlator}, {Livermore}, {Upton
  Sanderbeck}, {Dalla Vecchia}, \& {Khochfar}}]{Finkelstein2019}
{Finkelstein}, S.~L., {D'Aloisio}, A., {Paardekooper}, J.-P., {et~al.} 2019,
  \apj, 879, 36

\bibitem[{{Fletcher} {et~al.}(2019){Fletcher}, {Tang}, {Robertson}, {Nakajima},
  {Ellis}, {Stark}, \& {Inoue}}]{Fletcher2019}
{Fletcher}, T.~J., {Tang}, M., {Robertson}, B.~E., {et~al.} 2019, \apj, 878, 87

\bibitem[{{Flury} {et~al.}(2022{\natexlab{a}}){Flury}, {Jaskot}, {Ferguson},
  {Worseck}, {Makan}, {Chisholm}, {Saldana-Lopez}, {Schaerer}, {McCandliss},
  {Wang}, {Ford}, {Heckman}, {Ji}, {Giavalisco}, {Amorin}, {Atek}, {Blaizot},
  {Borthakur}, {Carr}, {Castellano}, {Cristiani}, {De Barros}, {Dickinson},
  {Finkelstein}, {Fleming}, {Fontanot}, {Garel}, {Grazian}, {Hayes}, {Henry},
  {Mauerhofer}, {Micheva}, {Oey}, {Ostlin}, {Papovich}, {Pentericci},
  {Ravindranath}, {Rosdahl}, {Rutkowski}, {Santini}, {Scarlata}, {Teplitz},
  {Thuan}, {Trebitsch}, {Vanzella}, {Verhamme}, \& {Xu}}]{Flury2022}
{Flury}, S.~R., {Jaskot}, A.~E., {Ferguson}, H.~C., {et~al.}
  2022{\natexlab{a}}, \apjs, 260, 1

\bibitem[{{Flury} {et~al.}(2022{\natexlab{b}}){Flury}, {Jaskot}, {Ferguson},
  {Worseck}, {Makan}, {Chisholm}, {Saldana-Lopez}, {Schaerer}, {McCandliss},
  {Xu}, {Wang}, {Oey}, {Ford}, {Heckman}, {Ji}, {Giavalisco}, {Amor{\'\i}n},
  {Atek}, {Blaizot}, {Borthakur}, {Carr}, {Castellano}, {De Barros},
  {Dickinson}, {Finkelstein}, {Fleming}, {Fontanot}, {Garel}, {Grazian},
  {Hayes}, {Henry}, {Mauerhofer}, {Micheva}, {Ostlin}, {Papovich},
  {Pentericci}, {Ravindranath}, {Rosdahl}, {Rutkowski}, {Santini}, {Scarlata},
  {Teplitz}, {Thuan}, {Trebitsch}, {Vanzella}, \& {Verhamme}}]{Flury2022b}
{Flury}, S.~R., {Jaskot}, A.~E., {Ferguson}, H.~C., {et~al.}
  2022{\natexlab{b}}, \apj, 930, 126

\bibitem[{{Gazagnes} {et~al.}(2020){Gazagnes}, {Chisholm}, {Schaerer},
  {Verhamme}, \& {Izotov}}]{Gazagnes2020}
{Gazagnes}, S., {Chisholm}, J., {Schaerer}, D., {Verhamme}, A., \& {Izotov}, Y.
  2020, \aap, 639, A85

\bibitem[{{Grazian} {et~al.}(2016){Grazian}, {Giallongo}, {Gerbasi}, {Fiore},
  {Fontana}, {Le F{\`e}vre}, {Pentericci}, {Vanzella}, {Zamorani}, {Cassata},
  {Garilli}, {Le Brun}, {Maccagni}, {Tasca}, {Thomas}, {Zucca}, {Amor{\'\i}n},
  {Bardelli}, {Cassar{\`a}}, {Castellano}, {Cimatti}, {Cucciati}, {Durkalec},
  {Giavalisco}, {Hathi}, {Ilbert}, {Lemaux}, {Paltani}, {Ribeiro}, {Schaerer},
  {Scodeggio}, {Sommariva}, {Talia}, {Tresse}, {Vergani}, {Bonchi}, {Boutsia},
  {Capak}, {Charlot}, {Contini}, {de la Torre}, {Dunlop}, {Fotopoulou},
  {Guaita}, {Koekemoer}, {L{\'o}pez-Sanjuan}, {Mellier}, {Merlin}, {Paris},
  {Pforr}, {Pilo}, {Santini}, {Scoville}, {Taniguchi}, \& {Wang}}]{grazian2016}
{Grazian}, A., {Giallongo}, E., {Gerbasi}, R., {et~al.} 2016, \aap, 585, A48

\bibitem[{{Grazian} {et~al.}(2017){Grazian}, {Giallongo}, {Paris}, {Boutsia},
  {Dickinson}, {Santini}, {Windhorst}, {Jansen}, {Cohen}, {Ashcraft},
  {Scarlata}, {Rutkowski}, {Vanzella}, {Cusano}, {Cristiani}, {Giavalisco},
  {Ferguson}, {Koekemoer}, {Grogin}, {Castellano}, {Fiore}, {Fontana},
  {Marchi}, {Pedichini}, {Pentericci}, {Amor{\'\i}n}, {Barro}, {Bonchi},
  {Bongiorno}, {Faber}, {Fumana}, {Galametz}, {Guaita}, {Kocevski}, {Merlin},
  {Nonino}, {O'Connell}, {Pilo}, {Ryan}, {Sani}, {Speziali}, {Testa}, {Weiner},
  \& {Yan}}]{Grazian2017}
{Grazian}, A., {Giallongo}, E., {Paris}, D., {et~al.} 2017, \aap, 602, A18

\bibitem[{Heckman {et~al.}(2011)Heckman, Borthakur, Overzier, Kauffmann,
  Basu-Zych, Leitherer, Sembach, Martin, Rich, Schiminovich, \&
  Seibert}]{Heckman2011}
Heckman, T.~M., Borthakur, S., Overzier, R., {et~al.} 2011, The Astrophysical
  Journal, 730, 5

\bibitem[{{Henry} {et~al.}(2015){Henry}, {Scarlata}, {Martin}, \&
  {Erb}}]{Henry2015}
{Henry}, A., {Scarlata}, C., {Martin}, C.~L., \& {Erb}, D. 2015, \apj, 809, 19

\bibitem[{{Horne}(1986)}]{Horne1986}
{Horne}, K. 1986, \pasp, 98, 609

\bibitem[{{Inoue} \& {Iwata}(2008)}]{inoue2008}
{Inoue}, A.~K. \& {Iwata}, I. 2008, \mnras, 387, 1681

\bibitem[{{Inoue} {et~al.}(2014){Inoue}, {Shimizu}, {Iwata}, \&
  {Tanaka}}]{inoue2014}
{Inoue}, A.~K., {Shimizu}, I., {Iwata}, I., \& {Tanaka}, M. 2014, \mnras, 442,
  1805

\bibitem[{{Izotov} {et~al.}(2017){Izotov}, {Guseva}, {Fricke}, {Henkel}, \&
  {Schaerer}}]{Izotov2017}
{Izotov}, Y.~I., {Guseva}, N.~G., {Fricke}, K.~J., {Henkel}, C., \& {Schaerer},
  D. 2017, \mnras, 467, 4118

\bibitem[{{Izotov} {et~al.}(2018){Izotov}, {Worseck}, {Schaerer}, {Guseva},
  {Thuan}, {Fricke}, \& {Orlitov{\'a}}}]{Izotov2018b}
{Izotov}, Y.~I., {Worseck}, G., {Schaerer}, D., {et~al.} 2018, \mnras, 478,
  4851

\bibitem[{{Jakobsen, P.} {et~al.}(2022){Jakobsen, P.}, {Ferruit, P.}, {Alves de
  Oliveira, C.}, {Arribas, S.}, {Bagnasco, G.}, {Barho, R.}, {Beck, T. L.},
  {Birkmann, S.}, {Böker, T.}, {Bunker, A. J.}, {Charlot, S.}, {de Jong, P.},
  {de Marchi, G.}, {Ehrenwinkler, R.}, {Falcolini, M.}, {Fels, R.}, {Franx,
  M.}, {Franz, D.}, {Funke, M.}, {Giardino, G.}, {Gnata, X.}, {Holota, W.},
  {Honnen, K.}, {Jensen, P. L.}, {Jentsch, M.}, {Johnson, T.}, {Jollet, D.},
  {Karl, H.}, {Kling, G.}, {Köhler, J.}, {Kolm, M.-G.}, {Kumari, N.}, {Lander,
  M. E.}, {Lemke, R.}, {López-Caniego, M.}, {Lützgendorf, N.}, {Maiolino,
  R.}, {Manjavacas, E.}, {Marston, A.}, {Maschmann, M.}, {Maurer, R.},
  {Messerschmidt, B.}, {Moseley, S. H.}, {Mosner, P.}, {Mott, D. B.},
  {Muzerolle, J.}, {Pirzkal, N.}, {Pittet, J.-F.}, {Plitzke, A.}, {Posselt,
  W.}, {Rapp, B.}, {Rauscher, B. J.}, {Rawle, T.}, {Rix, H.-W.}, {Rödel, A.},
  {Rumler, P.}, {Sabbi, E.}, {Salvignol, J.-C.}, {Schmid, T.}, {Sirianni, M.},
  {Smith, C.}, {Strada, P.}, {te Plate, M.}, {Valenti, J.}, {Wettemann, T.},
  {Wiehe, T.}, {Wiesmayer, M.}, {Willott, C. J.}, {Wright, R.}, {Zeidler, P.},
  \& {Zincke, C.}}]{jakobsen2022}
{Jakobsen, P.}, {Ferruit, P.}, {Alves de Oliveira, C.}, {et~al.} 2022, A\&A,
  661, A80

\bibitem[{{Jaskot} {et~al.}(2019){Jaskot}, {Dowd}, {Oey}, {Scarlata}, \&
  {McKinney}}]{Jaskot2019}
{Jaskot}, A.~E., {Dowd}, T., {Oey}, M.~S., {Scarlata}, C., \& {McKinney}, J.
  2019, \apj, 885, 96

\bibitem[{{Jaskot} \& {Oey}(2013)}]{Jaskot2013}
{Jaskot}, A.~E. \& {Oey}, M.~S. 2013, \apj, 766, 91

\bibitem[{Jaskot \& Oey(2013)}]{Jaskot_2013}
Jaskot, A.~E. \& Oey, M.~S. 2013, The Astrophysical Journal, 766, 91

\bibitem[{{Jaskot} {et~al.}(2024{\natexlab{a}}){Jaskot}, {Silveyra},
  {Plantinga}, {Flury}, {Hayes}, {Chisholm}, {Heckman}, {Pentericci},
  {Schaerer}, {Trebitsch}, {Verhamme}, {Carr}, {Ferguson}, {Ji}, {Giavalisco},
  {Henry}, {Marques-Chaves}, {{\"O}stlin}, {Saldana-Lopez}, {Scarlata},
  {Worseck}, \& {Xu}}]{Jaskot24a}
{Jaskot}, A.~E., {Silveyra}, A.~C., {Plantinga}, A., {et~al.}
  2024{\natexlab{a}}, \apj, 972, 92

\bibitem[{{Jaskot} {et~al.}(2024{\natexlab{b}}){Jaskot}, {Silveyra},
  {Plantinga}, {Flury}, {Hayes}, {Chisholm}, {Heckman}, {Pentericci},
  {Schaerer}, {Trebitsch}, {Verhamme}, {Carr}, {Ferguson}, {Ji}, {Giavalisco},
  {Henry}, {Marques-Chaves}, {{\"O}stlin}, {Saldana-Lopez}, {Scarlata},
  {Worseck}, \& {Xu}}]{Jaskot24b}
{Jaskot}, A.~E., {Silveyra}, A.~C., {Plantinga}, A., {et~al.}
  2024{\natexlab{b}}, \apj, 973, 111

\bibitem[{{Ji} {et~al.}(2020){Ji}, {Giavalisco}, {Vanzella}, {Siana},
  {Pentericci}, {Jaskot}, {Liu}, {Nonino}, {Ferguson}, {Castellano},
  {Mannucci}, {Schaerer}, {Fynbo}, {Papovich}, {Carnall}, {Amorin}, {Simons},
  {Hathi}, {Cullen}, \& {McLeod}}]{Ji2020}
{Ji}, Z., {Giavalisco}, M., {Vanzella}, E., {et~al.} 2020, \apj, 888, 109

\bibitem[{Jiang {et~al.}(2025)Jiang, Jiang, Sun, Liu, \& Fu}]{jiang2025}
Jiang, D., Jiang, L., Sun, S., Liu, W., \& Fu, S. 2025, arXiv preprint
  arXiv:2502.03683

\bibitem[{{Jung} {et~al.}(2024){Jung}, {Ferguson}, {Hayes}, {Henry}, {Jaskot},
  {Schaerer}, {Sharon}, {Amor{\'\i}n}, {Atek}, {Bayliss}, {Dahle},
  {Finkelstein}, {Grazian}, {Guaita}, {{\"O}stlin}, {Pentericci},
  {Ravindranath}, {Scarlata}, {Teplitz}, \& {Verhamme}}]{Jung2024}
{Jung}, I., {Ferguson}, H.~C., {Hayes}, M.~J., {et~al.} 2024, \apj, 971, 175

\bibitem[{{Jung} {et~al.}(2020){Jung}, {Finkelstein}, {Dickinson}, {Hutchison},
  {Larson}, {Papovich}, {Pentericci}, {Straughn}, {Guo}, {Malhotra}, {Rhoads},
  {Song}, {Tilvi}, \& {Wold}}]{jung2020}
{Jung}, I., {Finkelstein}, S.~L., {Dickinson}, M., {et~al.} 2020, \apj, 904,
  144

\bibitem[{{Katz} {et~al.}(2020){Katz}, {{\v{D}}urov{\v{c}}{\'\i}kov{\'a}},
  {Kimm}, {Rosdahl}, {Blaizot}, {Haehnelt}, {Devriendt}, {Slyz}, {Ellis}, \&
  {Laporte}}]{Katz2020}
{Katz}, H., {{\v{D}}urov{\v{c}}{\'\i}kov{\'a}}, D., {Kimm}, T., {et~al.} 2020,
  \mnras, 498, 164

\bibitem[{{Kerutt} {et~al.}(2024){Kerutt}, {Oesch}, {Wisotzki}, {Verhamme},
  {Atek}, {Herenz}, {Illingworth}, {Kusakabe}, {Matthee}, {Mauerhofer},
  {Montes}, {Naidu}, {Nelson}, {Reddy}, {Schaye}, {Simmonds}, {Urrutia}, \&
  {Vitte}}]{Kerutt2024}
{Kerutt}, J., {Oesch}, P.~A., {Wisotzki}, L., {et~al.} 2024, \aap, 684, A42

\bibitem[{Komarova {et~al.}(2024)Komarova, Oey, Hernandez, Adamo, Sirressi,
  Leitherer, Mas-Hesse, Östlin, Hodges-Kluck, Bik, Hayes, Jaskot, Kunth,
  Laursen, Melinder, \& Rivera-Thorsen}]{Komarova_2024}
Komarova, L., Oey, M.~S., Hernandez, S., {et~al.} 2024, The Astrophysical
  Journal, 967, 117

\bibitem[{{Kreilgaard} {et~al.}(2024){Kreilgaard}, {Mason}, {Cullen}, {Begley},
  \& {McLure}}]{kreilgaard24}
{Kreilgaard}, K.~C., {Mason}, C.~A., {Cullen}, F., {Begley}, R., \& {McLure},
  R.~J. 2024, \aap, 692, A57

\bibitem[{{Lin} {et~al.}(2024){Lin}, {Scarlata}, {Williams}, {Chen}, {Kelly},
  {Langeroodi}, {Hjorth}, {Chisholm}, {Koekemoer}, {Zitrin}, \&
  {Diego}}]{Lin2024}
{Lin}, Y.-H., {Scarlata}, C., {Williams}, H., {et~al.} 2024, \mnras, 527, 4173

\bibitem[{{Liu} {et~al.}(2023){Liu}, {Jiang}, {Windhorst}, {Guo}, \&
  {Zheng}}]{Liu2023}
{Liu}, Y., {Jiang}, L., {Windhorst}, R.~A., {Guo}, Y., \& {Zheng}, Z.-Y. 2023,
  \apj, 958, 22

\bibitem[{{Llerena} {et~al.}(2024){Llerena}, {Amor{\'\i}n}, {Pentericci},
  {Arrabal Haro}, {Backhaus}, {Bagley}, {Calabr{\`o}}, {Cleri}, {Davis},
  {Dickinson}, {Finkelstein}, {Gawiser}, {Grogin}, {Hathi}, {Hirschmann},
  {Kartaltepe}, {Koekemoer}, {McGrath}, {Mobasher}, {Napolitano}, {Papovich},
  {Pirzkal}, {Trump}, {Wilkins}, \& {Yung}}]{llerena2024}
{Llerena}, M., {Amor{\'\i}n}, R., {Pentericci}, L., {et~al.} 2024, \aap, 691,
  A59

\bibitem[{{Marchi} {et~al.}(2017){Marchi}, {Pentericci}, {Guaita}, {Ribeiro},
  {Castellano}, {Schaerer}, {Hathi}, {Lemaux}, {Grazian}, {Le F{\`e}vre},
  {Garilli}, {Maccagni}, {Amorin}, {Bardelli}, {Cassata}, {Fontana},
  {Koekemoer}, {Le Brun}, {Tasca}, {Thomas}, {Vanzella}, {Zamorani}, \&
  {Zucca}}]{Marchi2017}
{Marchi}, F., {Pentericci}, L., {Guaita}, L., {et~al.} 2017, \aap, 601, A73

\bibitem[{{Marchi} {et~al.}(2018){Marchi}, {Pentericci}, {Guaita}, {Schaerer},
  {Verhamme}, {Castellano}, {Ribeiro}, {Garilli}, {Le F{\`e}vre}, {Amorin},
  {Bardelli}, {Cassata}, {Durkalec}, {Grazian}, {Hathi}, {Lemaux}, {Maccagni},
  {Vanzella}, \& {Zucca}}]{Marchi2018}
{Marchi}, F., {Pentericci}, L., {Guaita}, L., {et~al.} 2018, \aap, 614, A11

\bibitem[{{Marques-Chaves} {et~al.}(2021){Marques-Chaves}, {Schaerer},
  {{\'A}lvarez-M{\'a}rquez}, {Colina}, {Dessauges-Zavadsky},
  {P{\'e}rez-Fournon}, {Saldana-Lopez}, \& {Verhamme}}]{Marques-Chaves2021}
{Marques-Chaves}, R., {Schaerer}, D., {{\'A}lvarez-M{\'a}rquez}, J., {et~al.}
  2021, \mnras, 507, 524

\bibitem[{{Marques-Chaves} {et~al.}(2022){Marques-Chaves}, {Schaerer},
  {{\'A}lvarez-M{\'a}rquez}, {Verhamme}, {Ceverino}, {Chisholm}, {Colina},
  {Dessauges-Zavadsky}, {P{\'e}rez-Fournon}, {Saldana-Lopez}, {Upadhyaya}, \&
  {Vanzella}}]{Marques-Chaves2022}
{Marques-Chaves}, R., {Schaerer}, D., {{\'A}lvarez-M{\'a}rquez}, J., {et~al.}
  2022, \mnras, 517, 2972

\bibitem[{{Mascia} {et~al.}(2024{\natexlab{a}}){Mascia}, {Pentericci},
  {Calabr{\`o}}, {Santini}, {Napolitano}, {Arrabal Haro}, {Castellano},
  {Dickinson}, {Ocvirk}, {Lewis}, {Amor{\'\i}n}, {Bagley}, {Bhatawdekar},
  {Cleri}, {Costantin}, {Dekel}, {Finkelstein}, {Fontana}, {Giavalisco},
  {Grogin}, {Hathi}, {Hirschmann}, {Holwerda}, {Jung}, {Kartaltepe},
  {Koekemoer}, {Lucas}, {Papovich}, {P{\'e}rez-Gonz{\'a}lez}, {Pirzkal},
  {Trump}, {Wilkins}, \& {Yung}}]{Mascia2024}
{Mascia}, S., {Pentericci}, L., {Calabr{\`o}}, A., {et~al.} 2024{\natexlab{a}},
  \aap, 685, A3

\bibitem[{{Mascia} {et~al.}(2023){Mascia}, {Pentericci}, {Calabr{\`o}}, {Treu},
  {Santini}, {Yang}, {Napolitano}, {Roberts-Borsani}, {Bergamini}, {Grillo},
  {Rosati}, {Vulcani}, {Castellano}, {Boyett}, {Fontana}, {Glazebrook},
  {Henry}, {Mason}, {Merlin}, {Morishita}, {Nanayakkara}, {Paris}, {Roy},
  {Williams}, {Wang}, {Brammer}, {Brada{\v{c}}}, {Chen}, {Kelly}, {Koekemoer},
  {Trenti}, \& {Windhorst}}]{Mascia2023}
{Mascia}, S., {Pentericci}, L., {Calabr{\`o}}, A., {et~al.} 2023, \aap, 672,
  A155

\bibitem[{{Mascia} {et~al.}(2024{\natexlab{b}}){Mascia}, {Roberts-Borsani},
  {Treu}, {Pentericci}, {Chen}, {Calabr{\`o}}, {Merlin}, {Paris}, {Santini},
  {Brammer}, {Henry}, {Kelly}, {Mason}, {Morishita}, {Nanayakkara}, {Roy},
  {Wang}, {Williams}, {Boyett}, {Brada{\v{c}}}, {Castellano}, {Glazebrook},
  {Jones}, {Napolitano}, {Vulcani}, {Watson}, \& {Yang}}]{Mascia2024_GLASSrel}
{Mascia}, S., {Roberts-Borsani}, G., {Treu}, T., {et~al.} 2024{\natexlab{b}},
  \aap, 690, A2

\bibitem[{{Mason} {et~al.}(2019){Mason}, {Fontana}, {Treu}, {Schmidt}, {Hoag},
  {Abramson}, {Amorin}, {Brada{\v{c}}}, {Guaita}, {Jones}, {Henry}, {Malkan},
  {Pentericci}, {Trenti}, \& {Vanzella}}]{mason2019}
{Mason}, C.~A., {Fontana}, A., {Treu}, T., {et~al.} 2019, \mnras, 485, 3947

\bibitem[{{Mauerhofer} {et~al.}(2021){Mauerhofer}, {Verhamme}, {Blaizot},
  {Garel}, {Kimm}, {Michel-Dansac}, \& {Rosdahl}}]{mauerhofer21}
{Mauerhofer}, V., {Verhamme}, A., {Blaizot}, J., {et~al.} 2021, \aap, 646, A80

\bibitem[{{Merlin} {et~al.}(2016){Merlin}, {Amor{\'\i}n}, {Castellano},
  {Fontana}, {Buitrago}, {Dunlop}, {Elbaz}, {Boucaud}, {Bourne}, {Boutsia},
  {Brammer}, {Bruce}, {Capak}, {Cappelluti}, {Ciesla}, {Comastri}, {Cullen},
  {Derriere}, {Faber}, {Ferguson}, {Giallongo}, {Grazian}, {Lotz},
  {Micha{\l}owski}, {Paris}, {Pentericci}, {Pilo}, {Santini}, {Schreiber},
  {Shu}, \& {Wang}}]{Merlin2016}
{Merlin}, E., {Amor{\'\i}n}, R., {Castellano}, M., {et~al.} 2016, \aap, 590,
  A30

\bibitem[{Meštrić {et~al.}(2021)Meštrić, Ryan-Weber, Cooke, Bassett,
  Prichard, \& Rafelski}]{mestric2021}
Meštrić, U., Ryan-Weber, E.~V., Cooke, J., {et~al.} 2021, Monthly Notices of
  the Royal Astronomical Society, 508, 4443

\bibitem[{Micheva {et~al.}(2017)Micheva, Iwata, Inoue, Matsuda, Yamada, \&
  Hayashino}]{Micheva2017}
Micheva, G., Iwata, I., Inoue, A.~K., {et~al.} 2017, Monthly Notices of the
  Royal Astronomical Society, 465, 316

\bibitem[{{Naidu} {et~al.}(2018){Naidu}, {Forrest}, {Oesch}, {Tran}, \&
  {Holden}}]{Naidu2018}
{Naidu}, R.~P., {Forrest}, B., {Oesch}, P.~A., {Tran}, K.-V.~H., \& {Holden},
  B.~P. 2018, \mnras, 478, 791

\bibitem[{{Nakajima} {et~al.}(2020){Nakajima}, {Ellis}, {Robertson}, {Tang}, \&
  {Stark}}]{Nakajima2020}
{Nakajima}, K., {Ellis}, R.~S., {Robertson}, B.~E., {Tang}, M., \& {Stark},
  D.~P. 2020, \apj, 889, 161

\bibitem[{{Nakajima} \& {Ouchi}(2014)}]{Nakajima2014}
{Nakajima}, K. \& {Ouchi}, M. 2014, \mnras, 442, 900

\bibitem[{{Napolitano} {et~al.}(2024){Napolitano}, {Pentericci}, {Santini},
  {Calabr{\`o}}, {Mascia}, {Llerena}, {Castellano}, {Dickinson}, {Finkelstein},
  {Amor{\'\i}n}, {Arrabal Haro}, {Bagley}, {Bhatawdekar}, {Cleri}, {Davis},
  {Gardner}, {Gawiser}, {Giavalisco}, {Hathi}, {Holwerda}, {Hu}, {Jung},
  {Kartaltepe}, {Koekemoer}, {Larson}, {Merlin}, {Mobasher}, {Papovich},
  {Park}, {Pirzkal}, {Trump}, {Wilkins}, \& {Yung}}]{Napolitano2024}
{Napolitano}, L., {Pentericci}, L., {Santini}, P., {et~al.} 2024, \aap, 688,
  A106

\bibitem[{{Oke} \& {Gunn}(1983)}]{Oke1983}
{Oke}, J.~B. \& {Gunn}, J.~E. 1983, \apj, 266, 713

\bibitem[{{Ouchi} {et~al.}(2020){Ouchi}, {Ono}, \& {Shibuya}}]{Ouchi_2020}
{Ouchi}, M., {Ono}, Y., \& {Shibuya}, T. 2020, \araa, 58, 617

\bibitem[{{Paardekooper} {et~al.}(2015){Paardekooper}, {Khochfar}, \& {Dalla
  Vecchia}}]{Paardekooper2015}
{Paardekooper}, J.-P., {Khochfar}, S., \& {Dalla Vecchia}, C. 2015, \mnras,
  451, 2544

\bibitem[{{Pahl} {et~al.}(2021){Pahl}, {Shapley}, {Steidel}, {Chen}, \&
  {Reddy}}]{Pahl2021}
{Pahl}, A.~J., {Shapley}, A., {Steidel}, C.~C., {Chen}, Y., \& {Reddy}, N.~A.
  2021, \mnras, 505, 2447

\bibitem[{{Pentericci} {et~al.}(2018){Pentericci}, {Vanzella}, {Castellano},
  {Fontana}, {De Barros}, {Grazian}, {Marchi}, {Bradac}, {Conselice},
  {Cristiani}, {Dickinson}, {Finkelstein}, {Giallongo}, {Guaita}, {Koekemoer},
  {Maiolino}, {Santini}, \& {Tilvi}}]{Pentericci_2018b}
{Pentericci}, L., {Vanzella}, E., {Castellano}, M., {et~al.} 2018, \aap, 619,
  A147

\bibitem[{{Planck Collaboration} {et~al.}(2020){Planck Collaboration},
  {Aghanim}, {Akrami}, {Ashdown}, {Aumont}, {Baccigalupi}, {Ballardini},
  {Banday}, {Barreiro}, {Bartolo}, {Basak}, {Battye}, {Benabed}, {Bernard},
  {Bersanelli}, {Bielewicz}, {Bock}, {Bond}, {Borrill}, {Bouchet}, {Boulanger},
  {Bucher}, {Burigana}, {Butler}, {Calabrese}, {Cardoso}, {Carron},
  {Challinor}, {Chiang}, {Chluba}, {Colombo}, {Combet}, {Contreras}, {Crill},
  {Cuttaia}, {de Bernardis}, {de Zotti}, {Delabrouille}, {Delouis}, {Di
  Valentino}, {Diego}, {Dor{\'e}}, {Douspis}, {Ducout}, {Dupac}, {Dusini},
  {Efstathiou}, {Elsner}, {En{\ss}lin}, {Eriksen}, {Fantaye}, {Farhang},
  {Fergusson}, {Fernandez-Cobos}, {Finelli}, {Forastieri}, {Frailis},
  {Fraisse}, {Franceschi}, {Frolov}, {Galeotta}, {Galli}, {Ganga},
  {G{\'e}nova-Santos}, {Gerbino}, {Ghosh}, {Gonz{\'a}lez-Nuevo}, {G{\'o}rski},
  {Gratton}, {Gruppuso}, {Gudmundsson}, {Hamann}, {Handley}, {Hansen},
  {Herranz}, {Hildebrandt}, {Hivon}, {Huang}, {Jaffe}, {Jones}, {Karakci},
  {Keih{\"a}nen}, {Keskitalo}, {Kiiveri}, {Kim}, {Kisner}, {Knox},
  {Krachmalnicoff}, {Kunz}, {Kurki-Suonio}, {Lagache}, {Lamarre}, {Lasenby},
  {Lattanzi}, {Lawrence}, {Le Jeune}, {Lemos}, {Lesgourgues}, {Levrier},
  {Lewis}, {Liguori}, {Lilje}, {Lilley}, {Lindholm}, {L{\'o}pez-Caniego},
  {Lubin}, {Ma}, {Mac{\'\i}as-P{\'e}rez}, {Maggio}, {Maino}, {Mandolesi},
  {Mangilli}, {Marcos-Caballero}, {Maris}, {Martin}, {Martinelli},
  {Mart{\'\i}nez-Gonz{\'a}lez}, {Matarrese}, {Mauri}, {McEwen}, {Meinhold},
  {Melchiorri}, {Mennella}, {Migliaccio}, {Millea}, {Mitra},
  {Miville-Desch{\^e}nes}, {Molinari}, {Montier}, {Morgante}, {Moss}, {Natoli},
  {N{\o}rgaard-Nielsen}, {Pagano}, {Paoletti}, {Partridge}, {Patanchon},
  {Peiris}, {Perrotta}, {Pettorino}, {Piacentini}, {Polastri}, {Polenta},
  {Puget}, {Rachen}, {Reinecke}, {Remazeilles}, {Renzi}, {Rocha}, {Rosset},
  {Roudier}, {Rubi{\~n}o-Mart{\'\i}n}, {Ruiz-Granados}, {Salvati}, {Sandri},
  {Savelainen}, {Scott}, {Shellard}, {Sirignano}, {Sirri}, {Spencer},
  {Sunyaev}, {Suur-Uski}, {Tauber}, {Tavagnacco}, {Tenti}, {Toffolatti},
  {Tomasi}, {Trombetti}, {Valenziano}, {Valiviita}, {Van Tent}, {Vibert},
  {Vielva}, {Villa}, {Vittorio}, {Wandelt}, {Wehus}, {White}, {White},
  {Zacchei}, \& {Zonca}}]{Planck2020}
{Planck Collaboration}, {Aghanim}, N., {Akrami}, Y., {et~al.} 2020, \aap, 641,
  A6

\bibitem[{{Price} {et~al.}(2025){Price}, {Bezanson}, {Labbe}, {Furtak}, {de
  Graaff}, {Greene}, {Kokorev}, {Setton}, {Suess}, {Brammer}, {Cutler}, {Leja},
  {Pan}, {Wang}, {Weaver}, {Whitaker}, {Atek}, {Burgasser}, {Chemerynska},
  {Dayal}, {Feldmann}, {F{\"o}rster Schreiber}, {Fudamoto}, {Fujimoto},
  {Glazebrook}, {Goulding}, {Khullar}, {Kriek}, {Marchesini}, {Maseda},
  {Miller}, {Muzzin}, {Nanayakkara}, {Nelson}, {Oesch}, {Shipley}, {Smit},
  {Taylor}, {Dokkum}, {Williams}, \& {Zitrin}}]{Price2025}
{Price}, S.~H., {Bezanson}, R., {Labbe}, I., {et~al.} 2025, \apj, 982, 51

\bibitem[{{Prieto-Lyon} {et~al.}(2023){Prieto-Lyon}, {Mason}, {Mascia},
  {Merlin}, {Roy}, {Henry}, {Roberts-Borsani}, {Morishita}, {Wang}, {Boyett},
  {Bolan}, {Bradac}, {Castellano}, {Mercurio}, {Nanayakkara}, {Paris},
  {Pentericci}, {Scarlata}, {Trenti}, {Treu}, \& {Vanzella}}]{Prieto-Lyon2023}
{Prieto-Lyon}, G., {Mason}, C., {Mascia}, S., {et~al.} 2023, \apj, 956, 136

\bibitem[{{Razoumov} \& {Sommer-Larsen}(2010)}]{Razoumov2010}
{Razoumov}, A.~O. \& {Sommer-Larsen}, J. 2010, \apj, 710, 1239

\bibitem[{{Reddy} {et~al.}(2016){Reddy}, {Steidel}, {Pettini},
  {Bogosavljevi{\'c}}, \& {Shapley}}]{Reddy2016}
{Reddy}, N.~A., {Steidel}, C.~C., {Pettini}, M., {Bogosavljevi{\'c}}, M., \&
  {Shapley}, A.~E. 2016, \apj, 828, 108

\bibitem[{{Richard} {et~al.}(2021){Richard}, {Claeyssens}, {Lagattuta},
  {Guaita}, {Bauer}, {Pello}, {Carton}, {Bacon}, {Soucail}, {Lyon}, {Kneib},
  {Mahler}, {Cl{\'e}ment}, {Mercier}, {Variu}, {Tamone}, {Ebeling}, {Schmidt},
  {Nanayakkara}, {Maseda}, {Weilbacher}, {Bouch{\'e}}, {Bouwens}, {Wisotzki},
  {de la Vieuville}, {Martinez}, \& {Patr{\'\i}cio}}]{richard2021}
{Richard}, J., {Claeyssens}, A., {Lagattuta}, D., {et~al.} 2021, \aap, 646, A83

\bibitem[{{Rivera-Thorsen} {et~al.}(2019){Rivera-Thorsen}, {Dahle}, {Chisholm},
  {Florian}, {Gronke}, {Rigby}, {Gladders}, {Mahler}, {Sharon}, \&
  {Bayliss}}]{Rivera-Thorsen2019}
{Rivera-Thorsen}, T.~E., {Dahle}, H., {Chisholm}, J., {et~al.} 2019, Science,
  366, 738

\bibitem[{{Rivera-Thorsen, T. E.} {et~al.}(2022){Rivera-Thorsen, T. E.},
  {Hayes, M.}, \& {Melinder, J.}}]{Rivera-Thorsen2022}
{Rivera-Thorsen, T. E.}, {Hayes, M.}, \& {Melinder, J.} 2022, A\&A, 666, A145

\bibitem[{{Roberts-Borsani} {et~al.}(2022){Roberts-Borsani}, {Morishita},
  {Treu}, {Brammer}, {Strait}, {Wang}, {Bradac}, {Acebron}, {Bergamini},
  {Boyett}, {Calabr{\'o}}, {Castellano}, {Fontana}, {Glazebrook}, {Grillo},
  {Henry}, {Jones}, {Malkan}, {Marchesini}, {Mascia}, {Mason}, {Mercurio},
  {Merlin}, {Nanayakkara}, {Pentericci}, {Rosati}, {Santini}, {Scarlata},
  {Trenti}, {Vanzella}, {Vulcani}, \& {Willott}}]{Roberts-Borsani2022b}
{Roberts-Borsani}, G., {Morishita}, T., {Treu}, T., {et~al.} 2022, \apjl, 938,
  L13

\bibitem[{{Robertson}(2022)}]{Robertson2022}
{Robertson}, B.~E. 2022, \araa, 60, 121

\bibitem[{{Roy} {et~al.}(2024){Roy}, {Heckman}, {Henry}, {Chisholm}, {Flury},
  {Leitherer}, {Hayes}, {Jaskot}, {Ji}, {Schaerer}, {Wang}, {Borthakur}, {Xu},
  \& {{\"O}stlin}}]{Roy2024}
{Roy}, N., {Heckman}, T., {Henry}, A., {et~al.} 2024, arXiv e-prints,
  arXiv:2410.13254

\bibitem[{{Roy} {et~al.}(2023){Roy}, {Henry}, {Treu}, {Jones}, {Prieto-Lyon},
  {Mason}, {Heckman}, {Nanayakkara}, {Pentericci}, {Mascia}, {Brada{\v{c}}},
  {Vanzella}, {Scarlata}, {Boyett}, {Trenti}, \& {Wang}}]{Roy2023}
{Roy}, N., {Henry}, A., {Treu}, T., {et~al.} 2023, \apjl, 952, L14

\bibitem[{Rutkowski {et~al.}(2017)Rutkowski, Scarlata, Henry, Hayes, Mehta,
  Hathi, Cohen, Windhorst, Koekemoer, Teplitz, Haardt, \&
  Siana}]{Rutkowski_2017}
Rutkowski, M.~J., Scarlata, C., Henry, A., {et~al.} 2017, The Astrophysical
  Journal Letters, 841, L27

\bibitem[{{Saldana-Lopez} {et~al.}(2022){Saldana-Lopez}, {Schaerer},
  {Chisholm}, {Flury}, {Jaskot}, {Worseck}, {Makan}, {Gazagnes}, {Mauerhofer},
  {Verhamme}, {Amor{\'\i}n}, {Ferguson}, {Giavalisco}, {Grazian}, {Hayes},
  {Heckman}, {Henry}, {Ji}, {Marques-Chaves}, {McCandliss}, {Oey},
  {{\"O}stlin}, {Pentericci}, {Thuan}, {Trebitsch}, {Vanzella}, \&
  {Xu}}]{Saldana-Lopez2022}
{Saldana-Lopez}, A., {Schaerer}, D., {Chisholm}, J., {et~al.} 2022, \aap, 663,
  A59

\bibitem[{{Sawant} {et~al.}(2021){Sawant}, {Pellegrini}, {Oey},
  {L{\'o}pez-Hern{\'a}ndez}, \& {Micheva}}]{Sawant2021}
{Sawant}, A.~N., {Pellegrini}, E.~W., {Oey}, M.~S., {L{\'o}pez-Hern{\'a}ndez},
  J., \& {Micheva}, G. 2021, \apj, 923, 78

\bibitem[{{Saxena} {et~al.}(2024){Saxena}, {Bunker}, {Jones}, {Stark},
  {Cameron}, {Witstok}, {Arribas}, {Baker}, {Baum}, {Bhatawdekar}, {Bowler},
  {Boyett}, {Carniani}, {Charlot}, {Chevallard}, {Curti}, {Curtis-Lake},
  {Eisenstein}, {Endsley}, {Hainline}, {Helton}, {Johnson}, {Kumari}, {Looser},
  {Maiolino}, {Rieke}, {Rix}, {Robertson}, {Sandles}, {Simmonds}, {Smit},
  {Tacchella}, {Williams}, {Willmer}, \& {Willott}}]{Saxena2022}
{Saxena}, A., {Bunker}, A.~J., {Jones}, G.~C., {et~al.} 2024, \aap, 684, A84

\bibitem[{{Siana} {et~al.}(2007){Siana}, {Teplitz}, {Colbert}, {Ferguson},
  {Dickinson}, {Brown}, {Conselice}, {de Mello}, {Gardner}, {Giavalisco}, \&
  {Menanteau}}]{Siana2007}
{Siana}, B., {Teplitz}, H.~I., {Colbert}, J., {et~al.} 2007, \apj, 668, 62

\bibitem[{Siana {et~al.}(2010)Siana, Teplitz, Ferguson, Brown, Giavalisco,
  Dickinson, Chary, de~Mello, Conselice, Bridge, Gardner, Colbert, \&
  Scarlata}]{Siana2010}
Siana, B., Teplitz, H.~I., Ferguson, H.~C., {et~al.} 2010, The Astrophysical
  Journal, 723, 241

\bibitem[{{Smail} {et~al.}(1997){Smail}, {Ivison}, \& {Blain}}]{Smail1997}
{Smail}, I., {Ivison}, R.~J., \& {Blain}, A.~W. 1997, \apjl, 490, L5

\bibitem[{{Steidel} {et~al.}(2018){Steidel}, {Bogosavljevi{\'c}}, {Shapley},
  {Reddy}, {Rudie}, {Pettini}, {Trainor}, \& {Strom}}]{Steidel2018}
{Steidel}, C.~C., {Bogosavljevi{\'c}}, M., {Shapley}, A.~E., {et~al.} 2018,
  \apj, 869, 123

\bibitem[{{Steidel} {et~al.}(2001){Steidel}, {Pettini}, \&
  {Adelberger}}]{steidel2001}
{Steidel}, C.~C., {Pettini}, M., \& {Adelberger}, K.~L. 2001, \apj, 546, 665

\bibitem[{{Trebitsch} {et~al.}(2017{\natexlab{a}}){Trebitsch}, {Blaizot},
  {Rosdahl}, {Devriendt}, \& {Slyz}}]{trebitsch17}
{Trebitsch}, M., {Blaizot}, J., {Rosdahl}, J., {Devriendt}, J., \& {Slyz}, A.
  2017{\natexlab{a}}, \mnras, 470, 224

\bibitem[{{Trebitsch} {et~al.}(2017{\natexlab{b}}){Trebitsch}, {Blaizot},
  {Rosdahl}, {Devriendt}, \& {Slyz}}]{Trebitsch2017}
{Trebitsch}, M., {Blaizot}, J., {Rosdahl}, J., {Devriendt}, J., \& {Slyz}, A.
  2017{\natexlab{b}}, \mnras, 470, 224

\bibitem[{{Treu} {et~al.}(2023){Treu}, {Calabr{\`o}}, {Castellano},
  {Leethochawalit}, {Merlin}, {Fontana}, {Yang}, {Morishita}, {Trenti},
  {Dressler}, {Mason}, {Paris}, {Pentericci}, {Roberts-Borsani}, {Vulcani},
  {Boyett}, {Bradac}, {Glazebrook}, {Jones}, {Marchesini}, {Mascia},
  {Nanayakkara}, {Santini}, {Strait}, {Vanzella}, \& {Wang}}]{Treu2022}
{Treu}, T., {Calabr{\`o}}, A., {Castellano}, M., {et~al.} 2023, \apjl, 942, L28

\bibitem[{{Vanzella} {et~al.}(2020){Vanzella}, {Caminha}, {Calura}, {Cupani},
  {Meneghetti}, {Castellano}, {Rosati}, {Mercurio}, {Sani}, {Grillo}, {Gilli},
  {Mignoli}, {Comastri}, {Nonino}, {Cristiani}, {Giavalisco}, \&
  {Caputi}}]{Vanzella2020}
{Vanzella}, E., {Caminha}, G.~B., {Calura}, F., {et~al.} 2020, \mnras, 491,
  1093

\bibitem[{{Vanzella} {et~al.}(2015){Vanzella}, {de Barros}, {Castellano},
  {Grazian}, {Inoue}, {Schaerer}, {Guaita}, {Zamorani}, {Giavalisco}, {Siana},
  {Pentericci}, {Giallongo}, {Fontana}, \& {Vignali}}]{vanzella_2015}
{Vanzella}, E., {de Barros}, S., {Castellano}, M., {et~al.} 2015, \aap, 576,
  A116

\bibitem[{{Vanzella} {et~al.}(2016){Vanzella}, {de Barros}, {Vasei}, {Alavi},
  {Giavalisco}, {Siana}, {Grazian}, {Hasinger}, {Suh}, {Cappelluti}, {Vito},
  {Amorin}, {Balestra}, {Brusa}, {Calura}, {Castellano}, {Comastri}, {Fontana},
  {Gilli}, {Mignoli}, {Pentericci}, {Vignali}, \& {Zamorani}}]{Vanzella2016}
{Vanzella}, E., {de Barros}, S., {Vasei}, K., {et~al.} 2016, \apj, 825, 41

\bibitem[{{Vanzella} {et~al.}(2012){Vanzella}, {Guo}, {Giavalisco}, {Grazian},
  {Castellano}, {Cristiani}, {Dickinson}, {Fontana}, {Nonino}, {Giallongo},
  {Pentericci}, {Galametz}, {Faber}, {Ferguson}, {Grogin}, {Koekemoer},
  {Newman}, \& {Siana}}]{Vanzella2012}
{Vanzella}, E., {Guo}, Y., {Giavalisco}, M., {et~al.} 2012, \apj, 751, 70

\bibitem[{{Vanzella} {et~al.}(2018){Vanzella}, {Nonino}, {Cupani},
  {Castellano}, {Sani}, {Mignoli}, {Calura}, {Meneghetti}, {Gilli}, {Comastri},
  {Mercurio}, {Caminha}, {Caputi}, {Rosati}, {Grillo}, {Cristiani}, {Balestra},
  {Fontana}, \& {Giavalisco}}]{Vanzella2018}
{Vanzella}, E., {Nonino}, M., {Cupani}, G., {et~al.} 2018, \mnras, 476, L15

\bibitem[{Vanzella {et~al.}(2010)Vanzella, Siana, Cristiani, \&
  Nonino}]{vanzella2010b}
Vanzella, E., Siana, B., Cristiani, S., \& Nonino, M. 2010, Monthly Notices of
  the Royal Astronomical Society, 404, 1672

\bibitem[{{Vanzella, E.} {et~al.}(2015){Vanzella, E.}, {de Barros, S.},
  {Castellano, M.}, {Grazian, A.}, {Inoue, A. K.}, {Schaerer, D.}, {Guaita,
  L.}, {Zamorani, G.}, {Giavalisco, M.}, {Siana, B.}, {Pentericci, L.},
  {Giallongo, E.}, {Fontana, A.}, \& {Vignali, C.}}]{Vanzella2015}
{Vanzella, E.}, {de Barros, S.}, {Castellano, M.}, {et~al.} 2015, A\&A, 576,
  A116

\bibitem[{{Verhamme} {et~al.}(2015){Verhamme}, {Orlitov{\'a}}, {Schaerer}, \&
  {Hayes}}]{Verhamme2015}
{Verhamme}, A., {Orlitov{\'a}}, I., {Schaerer}, D., \& {Hayes}, M. 2015, \aap,
  578, A7

\bibitem[{{Verhamme} {et~al.}(2017){Verhamme}, {Orlitov{\'a}}, {Schaerer},
  {Izotov}, {Worseck}, {Thuan}, \& {Guseva}}]{Verhamme2017}
{Verhamme}, A., {Orlitov{\'a}}, I., {Schaerer}, D., {et~al.} 2017, \aap, 597,
  A13

\bibitem[{Wang {et~al.}(2021)Wang, Heckman, Amorín, Borthakur, Chisholm,
  Ferguson, Flury, Giavalisco, Grazian, Hayes, Henry, Jaskot, Ji, Makan,
  McCandliss, Oey, Östlin, Saldana-Lopez, Schaerer, Thuan, Worseck, \&
  Xu}]{wang2021}
Wang, B., Heckman, T.~M., Amorín, R., {et~al.} 2021, The Astrophysical
  Journal, 916, 3

\bibitem[{Wang {et~al.}(2019)Wang, Heckman, Leitherer, Alexandroff, Borthakur,
  \& Overzier}]{wang2019}
Wang, B., Heckman, T.~M., Leitherer, C., {et~al.} 2019, The Astrophysical
  Journal, 885, 57

\bibitem[{{Wang} {et~al.}(2025){Wang}, {Teplitz}, {Smith}, {Windhorst},
  {Rafelski}, {Mehta}, {Alavi}, {Ji}, {Brammer}, {Colbert}, {Grogin}, {Hathi},
  {Koekemoer}, {Prichard}, {Scarlata}, {Sunnquist}, {Arrabal Haro},
  {Conselice}, {Gawiser}, {Guo}, {Hayes}, {Jansen}, {Lucas}, {O'Connell},
  {Robertson}, {Rutkowski}, {Siana}, {Vanzella}, {Ashcraft}, {Bagley},
  {Baronchelli}, {Barro}, {Blanche}, {Broussard}, {Carleton}, {Chartab},
  {Cheng}, {Codoreanu}, {Cohen}, {Dai}, {Darvish}, {Dav{\'e}}, {Degroot}, {de
  Mello}, {Dickinson}, {Emami}, {Ferguson}, {Ferreira}, {Finkelstein},
  {Finkelstein}, {Gardner}, {Gburek}, {Giavalisco}, {Grazian}, {Gronwall},
  {Hemmati}, {Howell}, {Iyer}, {Kaviraj}, {Kurczynski}, {Lazar}, {MacKenty},
  {Mantha}, {Martin}, {Martin}, {McCabe}, {Mobasher}, {Nedkova}, {Olsen},
  {Otteson}, {Ravindranath}, {Redshaw}, {Sattari}, {Soto}, {Yung}, {Zabelle},
  \& {UVCANDELS Team}}]{Wang2023}
{Wang}, X., {Teplitz}, H.~I., {Smith}, B.~M., {et~al.} 2025, \apj, 980, 74

\bibitem[{{Watson} {et~al.}(2025){Watson}, {Vulcani}, {Treu},
  {Roberts-Borsani}, {Dalmasso}, {He}, {Malkan}, {Morishita}, {Rojas Ruiz},
  {Zhang}, {Acharyya}, {Bergamini}, {Brada{\v{c}}}, {Fontana}, {Grillo},
  {Jones}, {Marchesini}, {Nanayakkara}, {Pentericci}, {Tubthong}, \&
  {Wang}}]{Watson2025}
{Watson}, P.~J., {Vulcani}, B., {Treu}, T., {et~al.} 2025, arXiv e-prints,
  arXiv:2504.00823

\bibitem[{Whitler {et~al.}(2025)Whitler, Stark, Topping, Robertson, Rieke,
  Hainline, Endsley, Chen, Baker, Bhatawdekar, {et~al.}}]{whitler2025}
Whitler, L., Stark, D.~P., Topping, M.~W., {et~al.} 2025, arXiv preprint
  arXiv:2501.00984

\bibitem[{{Wise} {et~al.}(2014){Wise}, {Demchenko}, {Halicek}, {Norman},
  {Turk}, {Abel}, \& {Smith}}]{Wise2014}
{Wise}, J.~H., {Demchenko}, V.~G., {Halicek}, M.~T., {et~al.} 2014, \mnras,
  442, 2560

\bibitem[{{Wyithe} \& {Loeb}(2013)}]{Wyithe2013}
{Wyithe}, J. S.~B. \& {Loeb}, A. 2013, \mnras, 428, 2741

\bibitem[{{Yamanaka} {et~al.}(2020){Yamanaka}, {Inoue}, {Yamada}, {Zackrisson},
  {Iwata}, {Micheva}, {Mawatari}, {Hashimoto}, \& {Kubo}}]{yamanaka2020}
{Yamanaka}, S., {Inoue}, A.~K., {Yamada}, T., {et~al.} 2020, \mnras, 498, 3095

\end{thebibliography}

\end{document}